\documentclass[hidelinks,12pt]{article}
\usepackage{setspace,amsmath,amssymb,amsthm,mathrsfs,sectsty,xcolor,graphicx,geometry,cite,hyperref,rotating,lscape}
\usepackage[square,sort&compress,comma,numbers]{natbib}
\hypersetup{
    colorlinks,
    linkcolor={red!50!black},
    citecolor={red!50!black},
    urlcolor={blue!50!green}
}

\begin{document}

\title{Studies on the Structure and Dynamics of Urban Bus Networks in Indian Cities}
\author{Atanu Chatterjee\\
Department of Civil Engineering\\
Indian Institute of Technology Madras\\
Chennai--600036, India}

\date{}

\maketitle

\begin{abstract}
\noindent In recent times, the domain of network science has become extremely useful in understanding the underlying structure of various real-world networks and to answer non-trivial questions regarding them. In this study, we rigourously analyze the statistical properties of the bus networks of six major Indian cities as graphs in \textit{L}- and \textit{P}-space, using tools from network science. Although public transport networks, such as airline and railway networks have been extensively studied, a comprehensive study on the structure and growth of bus networks is lacking. In India, where bus networks play an important role in day-to-day commutation, it is of significant interest to analyze their topological structure, and answer some of the basic questions on their evolution, growth, robustness and resiliency. We start from an empirical analysis of these networks, and determine their principle characteristics in terms of the complex network theory. The common features of small-world property and heavy tails in degree-distribution plots are observed in all the networks studied. Our analysis further reveals a wide spectrum of network topologies arising due to an interplay between preferential and random attachment of nodes. Unlike real-world networks, like the Internet, WWW and airline, which are virtual, bus networks are physically constrained in two-dimensional space by the underlying road networks. In order to understand the role of constraints in the evolution of these networks, we calculate their fractal dimensions that reveal a three-dimensional space-like evolution in a constrained two-dimensional plane. We also extend our study to understand the complex dynamical processes of epidemic outbreaks and information diffusion in these networks using SI and SIR models. Our findings, therefore throw light on the evolution and dynamics of such geographically and socio-economically constrained networks, which will help us in designing more efficient networks in the future.\\

\end{abstract}

\newpage
\tableofcontents
\newpage

\section{Review of previous works}
In this chapter, we give a brief sketch about the field of complexity science: the questions that led to the emergence of this new approach to understand real-world structures and their dynamics. We discuss in particular the emergence of the new field of research - \emph{Network Science}, and the various network models, their characteristic properties and their time-dependent dynamic behaviour. In relation to the theme of the thesis, we outline the significance of complex networks approach to understand public transportation infrastructures. Further, we review the literature in the area of structurally constrained network of bus routes in urban cities, which forms the core of the thesis. We end this section by elaborating on dynamical processes, like information percolation and phase-transition in networks in general, and epidemic spreading in transportation networks in particular.

\subsection{Emergence of complex network science}
\subsubsection{Introduction}
Nonlinearity is the essence of reality. The emergence of the field of complexity science is often attributed to the fact that real-world processes are indeterministic due to the presence of numerous variables and their nonlinear combinations. Real-world systems are also hard to comprehend because the agents constituting those systems show non-trivial interactions between them. In the context of these systems, it is always observed that the system as a whole is greater than its constituting parts~\cite{heylighen2001science, baranger2000chaos, gershenson2003can}. Therefore, these systems are complex not only because of their scale but also because of their functionality. The growing interest to understand the underlying machinery of these \emph{complex systems} has given rise to numerous mathematical and simulation techniques. Some of the widely used techniques include agent-based modelling, time series analysis, ant-colony optimization, cellular automata, nonlinear differential equations, information theory and network theory~\cite{shalizi2006methods, sayama2015introduction}. Amongst these existing mathematical methods, network theory in particular has been immensely successful in describing real world systems and processes in the recent times. 

\begin{figure}[hb!]
\begin{center}
\includegraphics[width=0.8\textwidth]{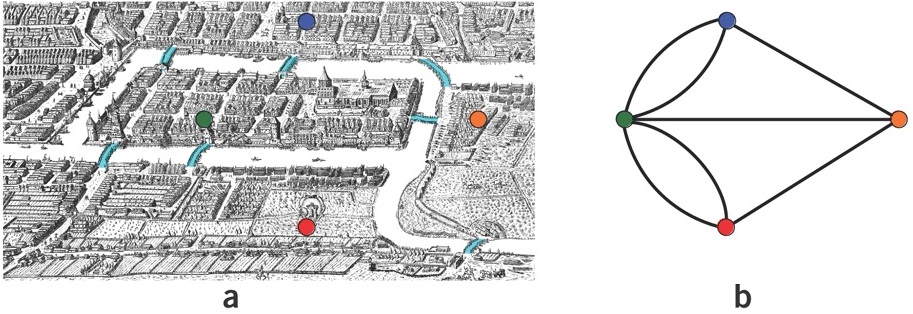}
\end{center}
\caption{Figure shows (a) the K{\"o}nigsberg bridge problem and (b) the corresponding graphical model~\cite{compeau2011apply}. The city of K{\"o}nigsberg in Prussia (now Kaliningrad, Russia) was set on both sides of the Pregel River, and included two large islands which were connected to each other and the mainland by seven bridges. The problem was to devise a walk through the city that would cross each bridge only once. The starting and the end points of the walk need not have been the same.}
\end{figure}

The underlying mathematical concepts of network theory or commonly, network science, is the theory of graphs. One of the oldest examples of using graph theory to analyze real-world problem dates back to as early as 1736, when the resolution to the famous K{\"o}nigsberg bridge problem by Euler actually laid the foundations of graph theory (see Figure 1). The abstract approach by Euler made the geographical intricacies present in the problem seem totally irrelevant. His although simple, yet abstract formulation of the problem in terms of vertices (nodes) and edges (links) provided an elegant approach to model real-world structures as graphs or networks. The first physical application of graph-theoretical ideas was discovered by Gustav Kirchhoff in 1845 for calculating the voltage and current in electric circuits. Since then, the use of graph theory as a modelling technique has found innumerable applications across diverse disciplines. One of the very important properties that these network models capture is the way different agents interact with each other in a connected system, which gives rise to non-trivial (emergent) properties in the system. 

Be it social ties or technological interactions, socio-economic infrastructures or interaction between biological entities, almost every thing can be modelled as a graph containing nodes and edges~\cite{albert2002statistical, strogatz2001exploring, bork2004protein, jeong2001lethality, stam2014modern, barabasi2004network, pastor2001epidemic, bollen2011twitter, jiang2010study, newman2003structure, deng2011exponential, chung2002average, cohen2003scale, PhysRevLett.94.018102}. Complex network science today has established itself as a mainstream field of research in the physical sciences. Although, the fundamental breakthroughs in network science are often attributed to statistical physicists and mathematicians, the origin of this field of study surprisingly lies in the social sciences. Two spectacular examples of studies that gave birth to network science as we know it today are that of Milgram's `small-world experiment' of 1967 and Granovetter's theory on the spread of information in social networks in 1973~\cite{milgram1967small, granovetter1973strength}. The results from Milgram's experiment helped us to understand how closely connected we are through our social ties whereas Granovetter's theory stressed on the importance of the individual connection that a node shares in the network. 

Both the observations led to an increased surge of interest in this new field of study by mathematicians and social scientists. With the advancement in information technology that enabled availability of large amounts of real-world data and improved computational resources, physicists eventually came into the picture in the early 1990's. It was during this time that the structure and topology of large-scale complex networks were being studied and fundamental questions like the nature of connections and in general, the statistical properties of the network's building blocks or the nodes were being asked. The field which initially branched out from the mathematical theory of graphs suddenly found its applications and similarities to statistical physics. No longer were the properties of the single node a question of interest. The group behaviour of the nodes - how they connect to each other - became a significant non-trivial question of academic pursuit.

\subsubsection{Network characteristics}
The field of network science has emerged out of the mathematical theory of graphs. Therefore, the underlying mathematical structure of the networks and its various characteristic properties use the language of Graph Theory for theoretical and computational purposes. Before we go into the details of the theoretical questions of interests (those we pointed out earlier), it is necessary to first mathematically describe a network and discuss certain characteristic features of these graphs, which are of interest in the particular context of the thesis. We will also observe later that these are the properties which help in differentiating between various network models. 

\begin{figure}[t]
\begin{center}
\includegraphics[width=0.8\textwidth]{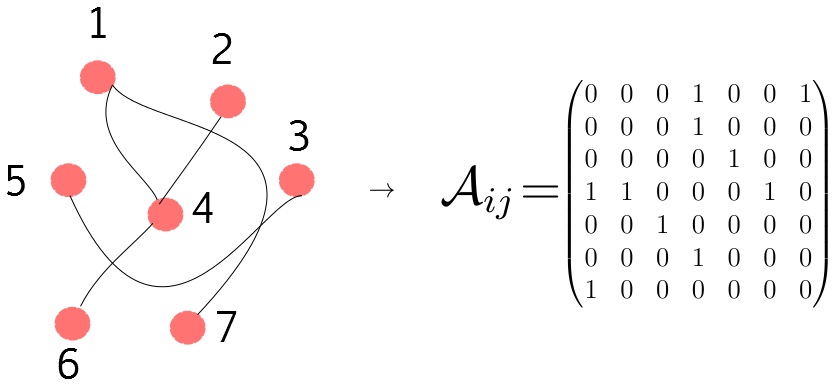}
\end{center}
\caption{Figure shows an undirected graph with seven nodes and the corresponding adjacency matrix. Note that the $7\times 7$ adjacency matrix is diagonally-symmetric.}
\end{figure}

We define a graph, $G = (N, L)$ where the set $N = (n_1 , n_2 , n_3 , ...)$ where each $n_i$ is a node, and the set $L = (l_{11} , l_{12} , l_{23} , ...)$ where each $l_{ij}$ is a link connecting the node pair $(n_i , n_j )$. The set of nodes belong to the n-dimensional Euclidean space, $\mathcal{R}^n$, and the set of links form the Cartesian product over $\mathcal{R}^n$. A graph can be directed or undirected. If a graph is directed, then $l_{ij}\neq l_{ji}$, whereas if a graph is undirected, we have $l_{ij}=l_{ji}$. Another important mathematical structure associated with graphs is the adjacency matrix, $\mathcal{A}_{ij}$. The adjacency matrix is a representation of the network as a $N\times N$ square matrix. Each element, $a_{ij}\in\mathcal{A}_{ij}$ either takes a value of $0, 1$ or $-1$ when the graph is unweighted. When $a_{ij}=0$ the link $l_{ij}$ does not exist, whereas $a_{ij}=1$ or $-1$ signifies the directionality of the link (with $i\rightarrow j$ in first case and $j\rightarrow i$ in the latter). If the graph in question is a weighted graph, then $a_{ij}$ can take any real value. The weights in the graph can represent diverse quantities, such as strength of connections (social network), frequency of interactions (call networks), travel times (transportation networks), distances (road networks), etc. For an undirected graph, the directionality of the links does not play any role ($l_{ij}=l_{ji}$), and the adjacency matrix takes on a diagonally-symmetric form. Also, note that $l_{ii}$ represents self-loops in the network (see Figure 2)~\cite{west2001introduction}. 

An important measure is the characteristic path length, $\textbf{l}_{ij}$ which is defined as the average number of nodes crossed along the shortest paths for all possible pairs of network nodes. The average distance from a certain vertex to every other vertex is given by $d_i = \Sigma_{i\neq j}\frac{d_{ij}}{|N(G)|-1}$. Then, $\textbf{l}_{ij}$ is calculated by taking the median of all the calculated $d_i$ $\forall i \in \mathcal{R}^n$. The characteristic path length helps in identifying whether a small-world phenomenon exists in the network or not. The significance of the small-world property is two-fold: one, it signifies very low characteristic path length, which simply means that given any pair of randomly chosen nodes, the number of `hops' needed to reach one from the other will be very low as compared to the network size; second, it also characterizes the change in the metric, $\textbf{l}_{ij}$ as the network size changes (the number of nodes present in a network is called its size)~\cite{albert2002statistical, strogatz2001exploring}. Milgram's experiment showed that on an average any randomly chosen pair of individuals are separated by six acquaintances (hence, the popular term: \emph{six-degrees of separation}). 

Another important network metric is the the clustering coefficient which measures the extent to which nodes tend to cluster in a network. The concept of clustering coefficient originated mainly from studying social networks, where a pair of individuals are connected by a link if they are friends. In social networks, close friends tend to have close-knit communities where each individual in the group knows every other individual. In Figure 3, we can observe that the maximum number of triangles possible for the blue node is three. In the first case, since every node is connected to every other node, the magnitude of clustering coefficient is one. Whereas, in the second and the last case when the dotted (red) links are removed, the possible number of triangles gets reduced to one and subsequently to zero. Hence, the magnitude of clustering coefficient in the second case is one-third and zero in the latter. The above example denotes the local clustering coefficient, which is in reference to a particular node in the network. The local clustering coefficient is given by:
\begin{equation}
C (i) =\frac{2|a_{ij}: (n_i , n_j)\in \mathcal{N}(i), a_{ij}\in \mathcal{A}_{ij}|}{k_i (k_i - 1)}
\end{equation} 
In the above expression, $k_i$ are the neighbours of the node $n_i$ and the neighbourhood, $\mathcal{N}(i)$, for a node, $n_i$ is defined as the set of its immediately connected neighbours, as $\mathcal{N}(i) = \{n_j : l_{ij}\in L \wedge l_{ji}\in L\}$. For the complete network, Watts and Strogatz defined a global clustering coefficient, $C =\Sigma_i C_i /N$~\cite{newman2003structure, watts1998collective}. Hence, the clustering coefficient measures the extent to which nodes tend to form close-knit groups in a network. A network with a small characteristic path length and a high clustering coefficient is called a small-world network.

\begin{figure}[t]
\begin{center}
\includegraphics[width=0.8\textwidth]{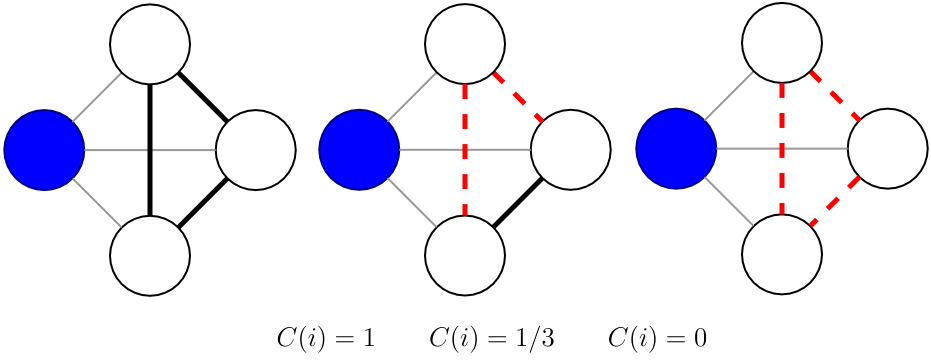}
\end{center}
\caption{Figure shows the clustering coefficient of the blue node ($C(i)$) in an undirected graph. In the first case, $C(i)=1$ as every node is connected to every other node. In the second case, the removal of two links (red, dotted) causes the number of triangles for the blue node to reduce to $1$, hence $C(i)=1/3$. In the last case, an additional removal of link causes the number of triangles for the blue node to reduce to zero, therefore $C(i)=0$.}
\end{figure}

Although the above metrics are crucial in identifying small-world phenomenon in a network, they do not help us in understanding the network topology. Understanding the network topology holds the key to several interesting and non-trivial questions. In order to answer questions about network robustness or node centrality, we need to look at the pattern by which the nodes are connected in the network. Before we go into the details of various degree-distribution patterns, we define the quantity - degree (of a node) in a network. The degree of a node is the number of neighbours to which it is directly connected to. In an undirected graph, the degree of a node is mathematically expressed as $k_i = \Sigma_i a_{ij}$. In a directed graph, the degree of a node will depend upon the directionality of the links which either emerge \emph{out} of the node or converge \emph{into} the node. Therefore, in a directed graph, a node will have an out-degree ($k_{out}$) and an in-degree ($k_{in}$), with $k_{in}=\Sigma_i a_{ij}$, where the index $i$ runs over all the positive $a_{ij}$ and vice-versa. When the network as a whole is studied, the significance of an individual node takes the backseat (unless we are interested in node-specific properties), and the way the degree of every single node is distributed throughout the network becomes a question of extreme interest. 

We define degree-distribution ($P(k_i)$) of a network as the probability that a node, $n_i$ has a degree of atleast $k_i$. The notion of degree-distribution holds a central role in network science. Networks that follow a similar degree-distribution law tend to show similar network characteristics. Therefore, the degree-distribution function can be used as a signature to differentiate between different network classes. Among all the possible degree-distribution laws, the ones that are most frequently encountered are the Poisson, exponential and power-law patterns. The Poissonion form of degree-distribution is given by:
\begin{equation}
P(k)=\exp(-\bar{k})\bar{k}^k /k! 
\end{equation}
Similarly, the exponential degree-distribution is given by:
\begin{equation}
P(k)\sim\exp(-k/\bar{k})\sim\exp(-\lambda k)
\end{equation}
And finally, the power-law degree distribution is given by:
\begin{equation}
P(k)\sim k^{-\gamma} \quad k\neq 0
\end{equation}
In the above expressions, $\bar{k}$ denotes the average node-degree, $\lambda$ and $\gamma$, the exponential and power-law degree exponents. Although all the distribution functions $P(k)$ (in eqn. 2, 3, 4) decay for large magnitudes of $k$, a special feature of the distributions in eqn. 2 and 3 is that they contain a typical scale. It is either the location of the maximum for the Poisson distribution, or the characteristic decay length for the exponential. On the contrary, the power-law distribution in eqn. 4 does not contain such a scale. Networks with a power-law degree-distribution are therefore called scale-free networks. When the degree-exponent, $\gamma < 2$, the average node degree diverges and when $\gamma < 3$, the standard deviation of the degree diverges~\cite{albert2002statistical}. Eqn. 2, 3 and 4 are probability density functions, therefore: $C\int P(k)dk=1$ for the continuous case and $C\Sigma_k P(k)=1$ for the discrete case. The constant, $C$ is known as the normalization constant. Clearly, the distribution in eqn. 4 diverges as $k\rightarrow 0$ so eqn. 4 cannot hold for all $k\geq 0$, \emph{i.e.}, there must be some lower-bound to the power-law behavior~\cite{clauset2009power}. We will denote this bound by $k_{min}$. Then, provided $\gamma > 1$, it is straightforward to calculate the normalizing constant, and we find that:
\begin{equation}
P(k) = \frac{\gamma - 1}{k_{min}}\Big(\frac{k}{k_{min}}\Big)^{-\gamma}
\end{equation}
For the discrete case, the distribution in eqn. 4 diverges when $k=0$, so there must be a lower bound $k_{min} > 0$ on the power-law behavior. On calculating the normalizing constant, we find that:
\begin{equation}
P(k) = \frac{k^{-\gamma}}{\zeta(\gamma, k_{min})}
\end{equation}
where the Hurwitz zeta function, $\zeta(\gamma, k_{min})$ is given by:
\begin{equation}
\zeta(\gamma, k_{min})=\Sigma_{n=0}^{\infty}\Big (n + k_{min}\Big)^{-\gamma}
\end{equation}

\begin{figure}[t]
\begin{center}
\includegraphics[width=0.25\textwidth]{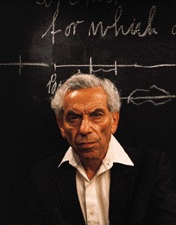}
\end{center}
\caption{Paul Erd{\"o}s (1913 - 1996): A highly connected mathematician with a prolific publication record. One of the earlier contributions made in the field of network science was the concept of random graphs in collaboration with Alfr{\'e}d R{\'e}nyi. The term \emph{Erd{\"o}s number} is used to identify the number of `hops' needed to connect the author of a paper with him~\cite{hoffman1987man}.}
\end{figure}

The Poisson distribution is strongly peaked about the mean $\bar{k}$, and has a tail that decays very rapidly as $1/k!$. This rapid decay is completely different from the heavy-tailed degree-distribution that is observed in many real-world complex networks. Real-world networks tend to show a degree heterogeneity, \emph{i.e.}, very a small fraction of nodes tend to hold majority of the connections in the network which are called as `hubs '. Instead of having a normally distributed degree in the network about the mean degree (Poissonian distribution function), real-world networks show tails which are heavily skewed towards the right. Also, it is hard to find real-world networks that show a perfect power-law or an exponential distribution, therefore majority of the networks tend to show a combination of them, such as power-law degree-distribution with exponential cut-offs or a combination of different power exponents. 

Whatever the case may be, it is important that we find proper explanations to these non-trivial characteristics of the real-world complex networks. In the following sections, we will elaborate on the important network models that will answer some of the interesting features of the networks that we saw earlier. Also, these are not the exhaustive list of network properties to be analyzed. We will see in the later sections that there are numerous other network metrics that describe both the global network properties and local nodal characteristics. 

\subsection{Complex network models}
In this section, we discuss the three prominent network models. Starting with the random graph model by Paul Erd{\"o}s (see Figure 4) and Alfred R{\'e}nyi in 1959 that laid the fundamental ideas of network science, we look into the small-world properties present in a network by Duncan Watts and Steven Strogatz in 1998. Finally, in order to understand the presence of heavy tails in degree-distribution patterns of real-world networks, we look into the scale-free network model proposed by Albert-L{\'a}szl{\'o} Barab{\'a}si and R{\'e}ka Albert also in 1998.

\subsubsection{Erd{\"o}s-R{\'e}nyi random graph}
The Erd{\"o}s-R{\'e}nyi (ER) model is either of the two closely related network models in Graph Theory for generating random graphs~\cite{renyi1959random, gilbert1959random}. In the model introduced by Erd{\"o}s and R{\'e}nyi, $G(n,l)$, the number of vertices ($n$) and the number of edges ($l$) are fixed, implying that all possible graph combinations are equally likely. Whereas in the model introduced by Gilbert, $G(n, p)$, each edge has a fixed probability $p$, of being present or absent, independently of the other edges. Therefore, in the $G(n, p)$ model, a graph is constructed by connecting nodes randomly with each edge having a fixed probability $p$ independent from every other edge. Therefore, all graphs with $n$ nodes and $l$ links have equal probability of $p^l (1-p)^{{n \choose 2}-l}$. As the parameter $p$ in this model increases from $0\rightarrow 1$, the model becomes more likely to include graphs with more edges and less likely to include graphs with fewer edges. In particular, when $p = 0.5$ all $2^{n\choose 2}$ graphs on $n$ vertices are chosen with equal probability. 

One of the interesting features of the ER random graph is the degree-distribution patterns of the nodes which follow the binomial form, $P(k) = {n-1\choose k} p^k (1-p)^{n-1-k}$ and approaches the Poissonian form as in eqn. 2 when $n\rightarrow\infty$ and $np=$ constant. Some of the important features about random graphs that Erd{\"o}s and R{\'e}nyi pointed out are~\cite{renyi1959random}:
\begin{itemize}
\item if $np < 1$, then a graph in $G(n, p)$ will almost surely have no connected components of size larger than $\mathcal{O}(log(n))$
\item if $np = 1$, then a graph in $G(n, p)$ will almost surely have a largest component whose size is of order $n^{2/3}$
\item if $np > 1$, then a graph in $G(n, p)$ will almost surely have a unique giant component containing a positive fraction of the vertices and no other component will contain more than $\mathcal{O}(log(n))$ vertices
\end{itemize}

Although these models ($G(n,l)$ and $G(n,p)$) can be used in the probabilistic method to prove the existence of graphs satisfying various properties, or to provide a rigorous definition of what it means for a property to hold for almost all graphs, they hardly capture the essence of the real-world network properties.

\begin{figure}[t]
\begin{center}
\includegraphics[width=1\textwidth]{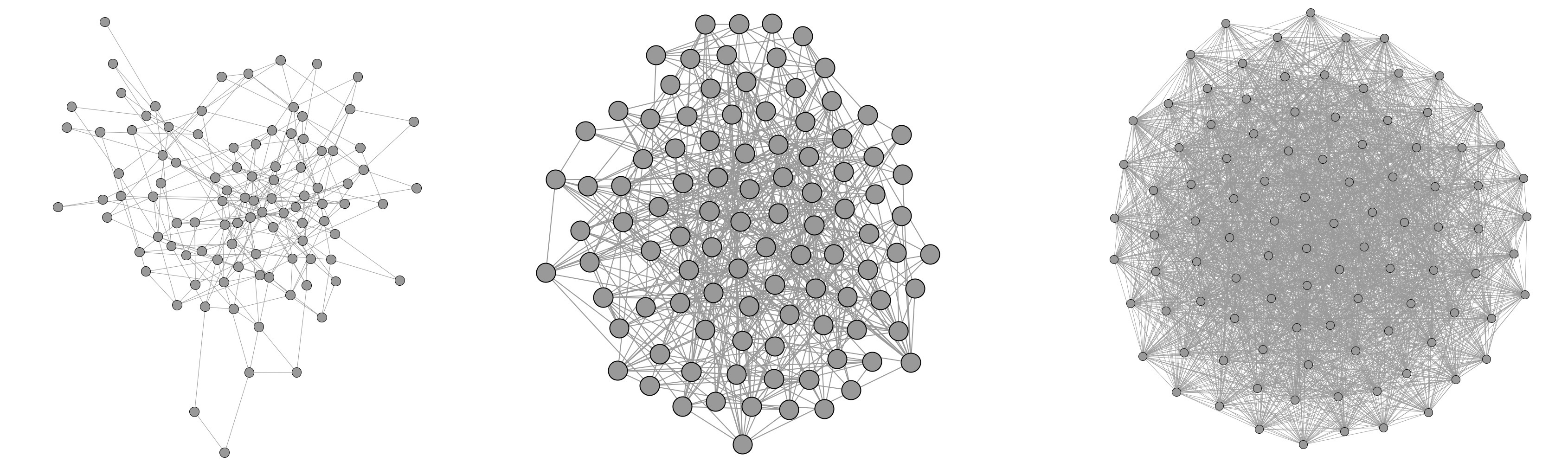}
\end{center}
\caption{Figure shows ER random graph, $G(n,p)$, with increasing wiring probability $p=0.05, 0.1$ and $0.5$ (left to right) and $n=100$. For $G(100,0.05)$ the number of links present $l=259$. Similarly for $G(100,0.1)\Rightarrow l=501$ and $G(100,0.5)\Rightarrow l=2490$. Observe how a ten-fold increase in probability causes the number of links to increase roughly by ten times.}
\end{figure}

\subsubsection{Watts-Strogatz small-world network}
The ER model provides a good mathematical description of graphs and the various properties associated with them. Although there may be instances when ER graphs show exceptionally small characteristic path lengths, they however do not show two important properties observed in many real-world networks:
\begin{itemize}
\item presence of clusters or triadic closures $\Rightarrow$ ER graphs have low clustering coefficient
\item presence of exceptionally high degree nodes or hubs $\Rightarrow$ degree-distributions in ER graphs converge to a Poisson form rather than a power-law pattern
\end{itemize}
In order to address the first of the above two limitations, Duncan Watts and Steven Strogatz devised a simple model that kept the characteristic path length small (similar to a random graph) but added the additional attribute of clustering. The model interpolated between a regular ring lattice and an ER random graph, and was able to explain the small-world phenomenon to some extent~\cite{watts1998collective}. Given a fixed number of nodes $N$, mean degree $\bar{k}$ (assumed to be an even integer), and a special parameter $p$, satisfying $0 \leq p \leq 1$ and $N>>\bar{k}>>\ln(N)>>1$, the algorithm constructs an undirected graph with $N$ nodes and $\frac{N\bar{k}}{2}$ edges in the following way (see Figure 6):
\begin{itemize}
\item construct a regular ring lattice, a graph with $N$ nodes each connected to $\bar{k}$ neighbors, $\bar{k}/2$ on either side, \emph{i.e.}, if the nodes are labeled $n_0 \ldots n_{N-1}$, there is an edge $(n_i, n_j)$ if and only if  $0 < |i - j|\mod (N-1-\frac{\bar{k}}{2}) \leq \frac{\bar{k}}{2}$
\item for every node $n_i=n_0,\dots, n_{N-1}$ take every edge $(n_i, n_j)$ with $i < j$, and rewire it with probability $p$. Rewiring is done by replacing $(n_i, n_j)$ with $(n_i, n_l)$ where $l$ is chosen with uniform probability from all possible values that avoid self-loops $(l \neq i)$ and link duplication (there is no edge $(n_i, n_{l'})$ with ${l'} = k$ at this point in the algorithm)
\end{itemize}

\begin{figure}[hb!]
\begin{center}
\includegraphics[width=0.75\textwidth]{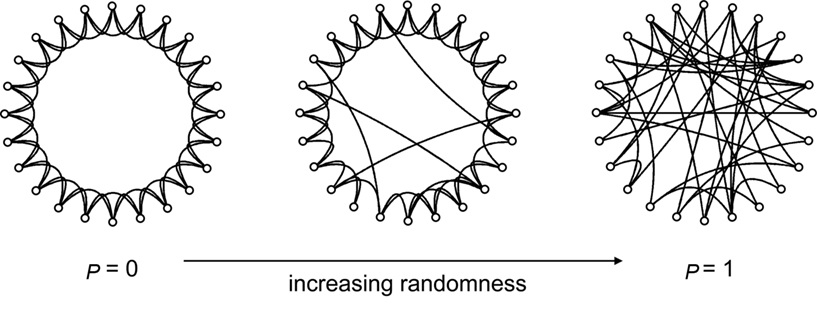}
\end{center}
\caption{Figure illustrates the Watts-Strogatz algorithm to generate small-world property in random graphs. As we can see above, the parameter $p$ plays a crucial role. When $p=0$, we have a regular ring lattice, whereas at $p =1$, we have a random graph, $G(n,p)$ with $n=N$ and $p=\frac{N\bar{k}}{{N\choose 2}}$~\cite{watts1998collective}.}
\end{figure}

We list out some network characteristics of the Watts-Strogatz small-world network model (WS). The degree distribution in the case of the ring lattice is the Dirac delta function centered at $\bar{k}$, which assumes the Poisson form in the limiting case of $\beta \rightarrow 1$ (eqn. 2), similar to the classical ER random graphs~\cite{barrat2000properties}. The degree distribution for $0<\beta <1$ can be written as:
\begin{equation}
P(k) = \Sigma_{n=0}^{f(k,\bar{k})} C^n_{\bar{k}/2} (1-p)^{n} p^{\bar{k}/2-n} \frac{(p\bar{k}/2)^{k-\bar{k}/2-n}}{(k-\bar{k}/2-n)!} \exp^{-p \bar{k}/2}
\end{equation}
where $k_i$ is the degree of the $i^{th}$ node, $k\geq\bar{k}/2$, and $f(k,\bar{k})=\min(k-\bar{k}/2,\bar{k}/2)$. The shape of the degree-distribution curve is similar to that of a random graph and has a pronounced peak at $k=\bar{k}$, which decays exponentially for large $|k-\bar{k}|$. The topology of the network is relatively homogeneous, and all nodes have more or less the same degree.

The additional feature that the WS model exhibits is the property of forming triadic closures or clusters. The clustering coefficient is given as the ratio of the number of triangles present among the nodes to the total number of connected triplets among them (see Figure 3). In terms of $p$,
\begin{equation}
C(p)\sim C(0)(1-p)^3 
\end{equation}
where, $C(0)$ is the clustering coefficient of a regular ring lattice given by $C(0)=\frac{3(\bar{k}-2)}{4(\bar{k}-1)}$ and $C(0)\rightarrow 3/4$ as $\bar{k}\rightarrow\infty$.

The WS model preserves the short characteristic path lengths exhibited by random graphs. For a regular ring lattice, the characteristic path length is given by $\textbf{l}_{ij}(0)=N/2\bar{k}$, which in the limiting case $p\rightarrow 1$ approaches to $\textbf{l}_{ij}=\frac{\ln(N)}{\ln(\bar{k})}$ or $\textbf{l}_{ij}\sim\ln(N)$ for a random graph. Thus, the number of hops required to visit all the nodes in a network scales as the logarithm of the network size, which still remains a very low number even though the network size is large (or grows with time). 

The small-world property has been reported in various real-world networks such as electric power grids, WWW, Internet, social-networks, protein-yeast (metabolite) interaction networks, citation networks, movie-actors collaboration networks \cite{barabasi1999emergence, albert2002statistical, albert1999internet, newman2003structure, watts1998collective, albert2004structural, bork2004protein, jeong2001lethality, easley2010networks, i2003least}. Although these networks show small characteristic path lengths and high clustering, they also exhibit a heavy-tailed degree-distribution that can not be explained either by the ER or the WS models. In order to understand the mechanism of growth and evolution of real-world networks, we analyze the Barab{\'a}si-Albert scale-free model in the following section. 

\subsubsection{Barab{\'a}si-Albert scale-free network}
Albert-L{\'a}szl{\'o} Barab{\'a}si, along with his doctoral student R{\'e}ka Albert published two important articles in the late 90's~\cite{barabasi1999emergence, albert1999internet}. In both the articles, they pointed out the peculiarity in the patterns of degree-distribution in real-world networks. They observed that the degree-distribution laws did not follow the Poissonian form nor the usually expected normal form. The density functions decayed slower than exponentials and log-normals, and showed a heavy-tail towards their end. Interestingly, their studies also confirmed the small-world property in these real-world datasets. 

In order to understand the growth and evolution of real-world networks exhibiting scale-free nature, Barab{\'a}si and Albert proposed an algorithm (BA model) that generates random networks with power-law degree-distribution patterns. The two salient features of the BA model are: \emph{growth} and \emph{preferential attachment}, both of which widely exist in real-world networks. The presence of preferential attachment allowed networks to exhibit degree-heterogeneity. A small fraction of nodes tend to attract more connections as compared to others, for example, highly influential people in a social network, important airports in an airline network or web pages, like Google. In a way, preferential attachment acted as a positive feedback loop for the network or initiated the `rich getting richer' phenomenon in the network. 

The earlier models failed to provide an explanation to this particularly observed phenomenon. Barab{\'a}si and Albert in their model incorporated this property and came up with the class of scale-free networks. The BA model, with growth and preferential attachment can be mathematically described using continuum theory, which calculates the time-dependence of the degree of a node~\cite{albert2002statistical}. Incorporating growth causes the network size to increase with time due to new incoming nodes. The preferential attachment is taken care by assigning a probability, $\Pi(k_i)$ for the new node to attach to an existing node in the network, such that $\Pi(k_i)=\frac{k_i}{\Sigma_{j=1}^{j=N-1}k_j}$. It is assumed that $k_i$ is a continuous real variable and the rate at which $k_i$ changes is proportional to $\Pi(k_i)$. Starting with a small number ($m_o$) of nodes, we add at every time-step a new node with $m$ edges. The real variable, $k_i$ satisfies the dynamical equation:
\begin{equation}
\frac{\partial k_i}{\partial t}=m\Pi(k_i)=m\frac{k_i}{\Sigma_{j=1}^{j=N-1}k_j}
\end{equation}
where $\Sigma_{j=1}^{j=N-1}k_j = 2mt-m$ leading to:
\begin{equation}
\frac{\partial k_i}{\partial t}=\frac{k_i}{2t}
\end{equation}
The solution to the above equation with the initial condition that every node introduced at time $t_i$ has $m$ edges $\Rightarrow k_i (t_i)=m$ is given by $k_i (t)=m\Big(\frac{t}{t_i}\Big)^\beta$, with $\beta=1/2$. Using eqn. 11, the probability that a node has a degree $k_i (t)$ smaller than $k$, $P(k_i (t)<k)$ can be written as: 
\begin{equation}
P(k_i (t)<k)=P\Big (t_i > \frac{m^{\frac{1}{\beta}}t}{k^{\frac{1}{\beta}}}\Big)
\end{equation}
Since the nodes are added at equal time intervals, the parameter $t_i$ has a constant probability density, $P(t_i)=\frac{1}{m_o + t}$. Substituting the value of $P(t_i)$ into eqn. 12, we obtain:
\begin{equation}
P\Big (t_i > \frac{m^{\frac{1}{\beta}}t}{k^{\frac{1}{\beta}}}\Big) = 1 - \frac{m^{\frac{1}{\beta}}t}{k^{\frac{1}{\beta}}(m_o + t)}
\end{equation}
The degree-distribution $P(k)$ can be obtained by:
\begin{equation}
P(k)=\frac{\partial P(k_i (t)<k)}{\partial k}=\frac{2m^{\frac{1}{\beta}}t}{k^{\frac{1}{\beta}+1}(m_o + t)}
\end{equation}
The above equation can be written in the asymptotic ($t\rightarrow\infty$) limit as $P(k)\sim 2m^{\frac{1}{\beta}}k^{-\gamma}$, with $\gamma = \frac{1}{\beta} + 1 = 3$. The above result is independent of the number of links ($m$) an incoming node has. Therefore, BA model produces scale-free networks with degree-distribution following pure power-laws with degree exponent, $\gamma = 3$ (see Figure 7).

Although BA model generates networks obeying pure power-law degree-distributions, real-world networks on the contrary mostly show a heavy tail. The reason for this can be attributed to several factors, such as non-linear preferential attachment, node aging, node's competitiveness to acquire incoming links, or presence of exponential cut-offs~\cite{albert2002statistical}. Even though real-world networks may not follow strict power-laws, the presence of preferential attachment is extremely crucial. Without the presence of the preferential attachment, rule networks cannot exhibit heavy-tailed degree-distribution patterns because the preferential attachment rule acts as a positive reinforcement in the network. BA models, without growth have also been proposed. However, network models without the attachment rule will result into random graphs~\cite{xie2008scale}. The characteristic path length for BA networks is given as $\textbf{l}_{ij}\sim\frac{\log N}{\log\log k}$. For $2<\gamma<3$, the characteristic path length has been found to decay as $\textbf{l}_{ij}\sim\ln\ln N$, thus making the topology of scale-free networks ultra-small in nature~\cite{cohen2003scale}.

\begin{figure}[t]
\begin{center}
\includegraphics[width=1\textwidth]{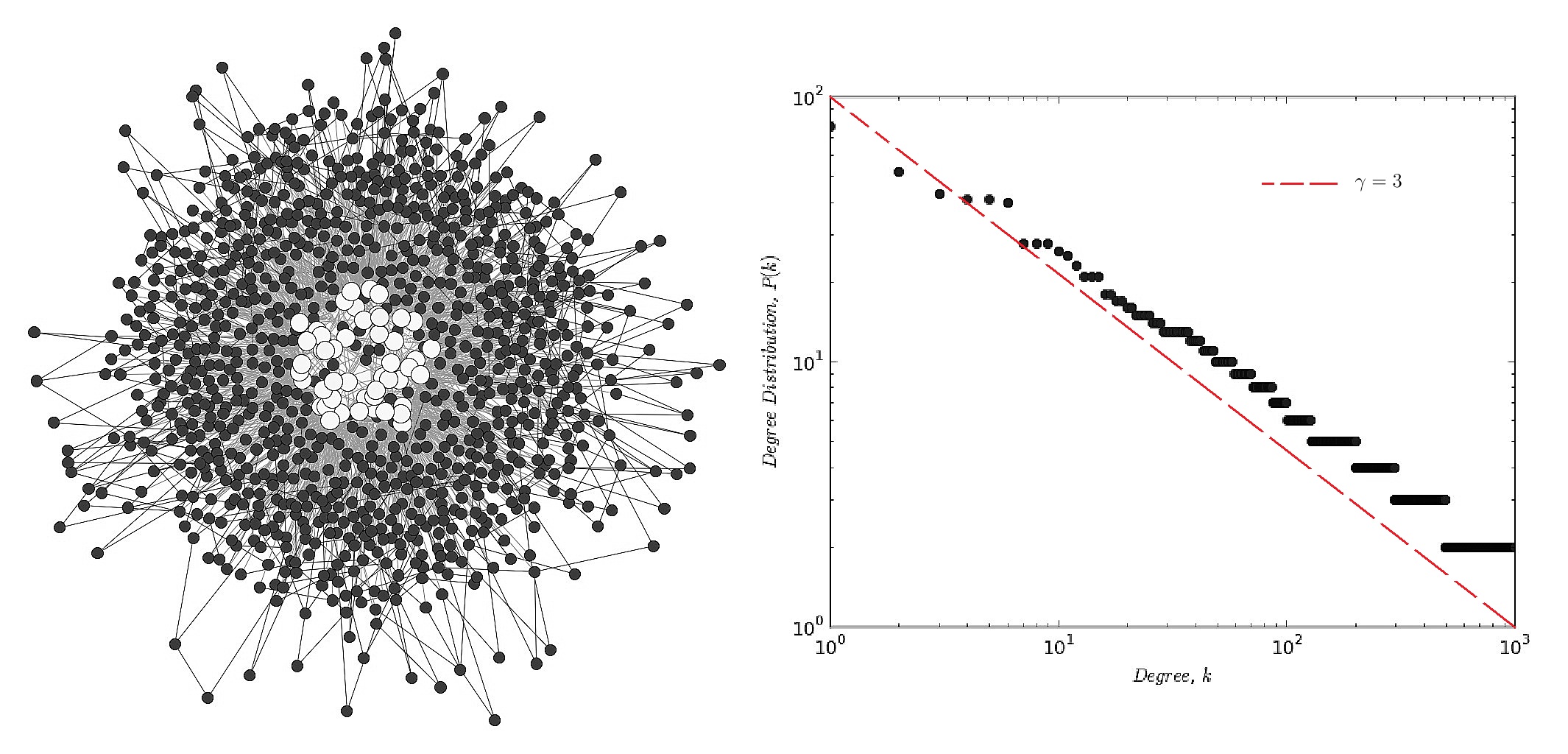}
\end{center}
\caption{The left panel shows a scale-free network, with $n = 1000$ and $l = 3187$. The node size reflects the node degree, $i.e.$, nodes which have bigger size have higher degrees and vice versa. In the right panel, we plot the degree-distribution of the network on a double logarithmic scale. The degree-distribution follows a power-law pattern, $P(k)\sim k^{\gamma}$, with $\gamma = 3$. The parameter, $\gamma$ is calculated from the slope of the degree-distribution plot in the right panel (red, dotted).}
\end{figure}

\subsection{Complex networks approach to transportation science}
The availability of huge amounts of real-time data and sudden surge of interest in the field of networks have attracted numerous researchers working in applied sciences as well. Particular in this context are those civil engineers whose research is primarily focused in the domain of transportation networks. Although the study of transportation networks have a long history, the questions of interests have significantly changed after the introduction of network science techniques. The traditional network flow formulation (in  transportation networks) has answered many interesting engineering questions related to optimality of cost, maximality of flows and the classical shortest path determination~\cite{ahuja1988network, bertsimas2003robust}. Questions, those related to the topological structure of the network, such as the presence of small-world property or heavy-tailed degree-distribution patterns, or questions which are primarily concerned with the inter-nodal connectivity and static (dynamic) evolution of the network which the traditional formulation failed to address, found answers in the network science domain. 

One particular type of transportation network that found widespread interest in the network science community is that of the public transit network or PTN. A topological drawback of PTNs is that they are structurally constrained in a two-dimensional space as compared to other networks, such as the Internet, social networks or the airline networks. Airline and metro-networks, specifically have been reported to show scale-free degree distribution patterns, whereas degree-distribution in bus and railway networks tend more towards exponential forms (or to power-law patterns with larger magnitudes of $\gamma$). Yet, the above mentioned properties - small-world phenomenon and scale-free topology - have been reported in them as well \cite{derrible2009network, zhang2014analysis, sen2003small, bagler2008analysis, von2009public, woolley2011complexity, guimera2005worldwide, sienkiewicz2005statistical, angeloudis2006large}. In some studies, specific subsets of PTNs were analyzed, for example, the Boston subway network, the Vienna subway network, or bus networks of three cities in China~\cite{marchiori2000harmony, latora2001efficient, latora2002boston, seaton2004stations, xu2007scaling}. It is important to note that each separate type of public transport (bus, subway, trams or mono-rail network) is not a closed system: these are only sub-networks of a much wider city transport system. An interesting observation was reported when network characteristics obtained while separately analyzing subway networks were compared to those of the network combination of subway and buses~\cite{latora2001efficient, latora2002boston}. Statistical analysis of complete PTNs for cities, such as Berlin, Paris, D{\"u}sseldorf and 22 Polish cities showed that power-law degree-distribution pattern is a common feature~\cite{von2009public, sienkiewicz2005statistical}. They also concluded that the degree-distribution patterns for PTNs can have both exponential as well as power-law forms, depending upon the topology of the network representation chosen and presence of geographical constraints. 

The small-world phenomenon in transportation networks makes sense as transportation facilities in a city are planned to provide maximum convenience to its people by allowing them to travel between places in minimum possible time. Most transportation networks are pre-planned networks, where the initial design of the network decides the presence of hubs. Also, the size of transportation networks is not as large compared to social-networks or the Internet, and are subjected to geographical as well as socio-economical constraints. The reason for the contrasting behaviour of the airline and metro networks from bus and rail networks can be attributed to the two following observations: (\textbf{i}) airline-networks (like the Internet or social-networks) are not bounded by geographical constraints, and (\textbf{ii}) metro-networks are \emph{local} often catering to a part of the city, whereas bus and railway-networks are \emph{global} as they are spread throughout the entire state and sometimes across the entire country. Specific to Indian scenarios, exhaustive studies on public transit networks as a whole are yet to be conducted. Previous works, have shown that the pattern of nodal connectivity of the Indian Railway Network (IRN) drastically differs from that of the Airport Network of India (ANI), while the nature of Indian bus networks still remains an unsolved problem~\cite{sen2003small, bagler2008analysis}.

\subsection{Studies on bus transport networks}
The central theme of the thesis lies around the topological structure and dynamics of bus transport networks or BTNs in India. The presence of comparatively less number of studies on BTNs among other modes of transportation is one of the fundamental reasons that their characteristic properties and their topological structures are yet inconclusive. In this section, we present the currently available literature on the studies pertaining to BTNs in specific.

Analysis of the statistical properties of BTNs in China have revealed scale-free degree-distribution patterns and small-world properties in them. The presence of nontrivial clustering, \emph{i.e.}, the variation of clustering with degree, indicated a hierarchical and modular structure in those BTNs. Weighted analysis of the network revealed a heavy-tailed power law with the strength (weighted-degree) and degree showing a linear dependency~\cite{xu2007scaling}. In another study, the BTNs of four major cities of China, namely, Hangzhou, Nanjing, Beijing and Shanghai were analyzed using $P$-space topology. The degree-distribution was reported to follow an exponential form, indicating a tendency for random attachment of the nodes. The authors also evaluated two new statistical properties of the BTNs: the distribution of number of stops in a bus route ($S$), and the number of bus routes a stop joins ($R$). While the former had an exponential functional form, the latter had an asymmetric unimodal functional form~\cite{chen2007study}. In a separate study, the urban public bus networks of two Chinese cities, Beijing and Chengdu were analyzed. Their analysis revealed small world characteristics and a scale-free topology, although with exceptionally high values for the degree-exponent, $\gamma$. The presence of more hubs in the Beijing network yielded a smaller $\gamma$ as compared to Chengdu; while both showed large clustering coefficients and small characteristic path lengths. The location of bus stops in a similar fashion in both the cities have led to a hierarchical structure, which is denoted by a power-law behaviour (with nearly same exponents) between the degree strength (characterizing the passenger flows) and clustering coefficient~\cite{ma2011power, ding2009study}. 

In a recent study, the combination of rail (RTNs) and bus transportation systems (BTNs) in Singapore were studied with respect to their topological as well as dynamical perspectives. The stations in RTNs had high average degree indicating high connectivity amongst them, while the BTNs had a small average degree. Both networks had an exponential degree-distribution indicative of randomly evolved connectivity. Strength-distribution of the nodes (weighted degree-distribution) however showed scale-free topology for both networks, indicating the existence of high traffic hubs. Both BTNs and RTNs exhibited small-world characteristics. The BTN in particular had a hierarchical star like topology. Degree-assortativity ($r$), which measures the inter-connectivity between hubs, revealed RTNs to be slightly disassortative, while the BTNs displayed strong disassortative nature~\cite{soh2010weighted}. With the availability of geo-located data, an extended space (ES) model with information on geographical location of bus stops and routes was recently used to analyze the spatial characteristics of BTNs in China. The $ES$ model consisted of directed weighted variations of the $L$- and $P$-space networks, designated as ES-$L$ and ES-$P$ networks respectively. Often, two bus stops which are geographically close to each other may not have any direct bus route link between them. However, such stops are at walkable distances from each other. These are defined as short-distance station pairs or SSPs. The SSPs greatly influence the BTNs by reducing the transfer times as well as the number of bus routes. The symmetry-weighted ESW network model stored information of the SSPs. The average clustering coefficient of the ESW network was considerably large, denoting a nearly circular location of the SSPs around a station. Majority of the route sections in the bus routes were short, while a few route sections connecting cities' downtowns and satellite towns or special purpose routes were long, leading to a power-law edge length distribution of the ES-$L$ network~\cite{yang2014study}. 

It may seem at first that the complexity of a bus transportation network is much lesser than that of other large-scale networks, however it is the nature of growth and the penetrative effect of these networks that makes them not only complex, but interesting and worthwhile to investigate.

\subsection{Dynamical processes on networks}
In the earlier sections, we discussed about the topological characteristics of the networks in detail. In order to get a deeper insight into the network characteristics, it is important to look at the dynamical process on networks as well. In this section, we discuss the two dynamical processes on networks in detail: network diffusion (SI model) and network contagion (SIR model). The spatial characteristics of the networks along with numerical simulations of the dynamical processes will help us understand the in-depth significance of various network metrics.

\subsubsection{Epidemic spreading in complex networks}
The earliest accounts of mathematical modelling to capture the spread of diseases dates back to as early as the $17^{th}$ century. Bernoulli used mathematical equations to defend his stand on vaccination against the outbreak of smallpox. Works following Bernoulli\rq{s} earliest formulation of epidemic modelling helped in understanding germ theory in detail. However, this was not until the works of McKendrick and Kermack, which first proposed a deterministic model that predicted epidemic outbreaks very similar to the ones that were recorded during those times~\cite{kermack1927contribution}. Since then our understanding of the mathematical models in epidemology has evolved over the years, the accounts of which can be found in the extensive works of Anderson and May~\cite{anderson1992infectious}. All the above formulations focused on modelling epidemics over a set of population in which uniform ties between agents were assumed \emph{a priori}. Contrary to this, the field of network science asserts the fact that ties, their strength and types in a system or a population, are not uniform and they play a significant role to describe the system's dynamics~\cite{albert2002statistical, granovetter1973strength, watts1998collective}. Also, a network model is not exhaustive to the study of population, rather it is a universal framework which can be used to understand numerous complex systems in general. 

Over the years, the term \emph{epidemic modelling} has evolved into a common metaphor for a wide array of dynamical processes on these networks. Various complex phenomena such as percolation, the spreading of blackouts to the spreading of memes, ideas and opinions in a social network have been modelled under the common framework of epidemic modelling~\cite{vespignani2009predicting}. In this context, transportation networks play a vital role due to their widespread outreach across cities, countries and continents~\cite{pastor2001epidemic}. \lq\lq{S}hould people be worried about getting Ebola on the subway$?$\rq\rq{,} was one of the numerous similar headlines that made the front pages of the newspapers around the world during the 2014 Ebola scare~\cite{paper}. In this particular incident however, nobody was infected because the subject did not show symptoms of Ebola while using public transportation. Therefore not only airline networks, that can transmit pathogens across continents, even modes of public transport operating within cities, such as buses and subways, pose a serious threat as well as a source of panic during desperate times. Although epidemic spreading in airline networks have been studied extensively, similar studies on bus networks are relatively rare~\cite{albert2002statistical, pastor2001epidemic}. Epidemological models have been simulated on bus network datasets; however, the results were only used to validate the numerical models. Also, a recent study on city-wide Integrated Travel Networks (ITNs), has found both the traveling speed and frequency to be important factors in epidemic spreading~\cite{ruan2015integrated}. Thus, the effect of network structure and constraints in epidemic spreading is yet to be studied in these networks.

\subsubsection{The SI model: network diffusion}
The SI model is the most basic representation of an epidemic spreading model that captures diffusion in complex networks. In this model, there are two states that an agent or a node can exist in: S (susceptible) or I (infected). The SI model describes the status of individuals or agents switching from susceptible to infected at every instant of time. It is assumed that the population is homogeneous and closed, \emph{i.e.}, no new entity is either created due to birth or removed due to death, and also, no new entity enters the system, thus preserving homogeneous mixing in the system. The SI model also implies that each individual has the same probability to transfer disease, innovation or information to its neighbors. Thus, the SI model helps to capture the diffusion or percolation process in the entire network. The SI model is formulated using the following differential equation. Since an agent in the entire population can either be in state S or I, 
\begin{equation}
S + I = 1
\end{equation}
The SI model is governed by a single parameter, $\beta$, the infection transmission rate or simply, the infection rate. The growth in the number of agents in either of the sates is given by:
\begin{equation}
\frac{dS}{dt}=\frac{dI}{dt}=-\beta SI
\end{equation}
Substituting the value of S from eqn. 15 to eqn. 16, we get the following differential equation describing the growth rate of I:
\begin{equation}
\frac{dI}{dt}=-\beta (1-I)I
\end{equation}
The solution of the above equation with the initial condition at $t=0$, $I = I_0$ is given by the logistic form:
\begin{equation}
I = \Big (1 + \exp(-\beta t)\Big (\frac{1-I_0}{I_0}\Big )\Big )^{-1}
\end{equation}

\subsubsection{The SIR model: network contagion}
Contrary to the SI model, the agents in SIR model have access to three states: S (susceptible), I (infected) and R (recovery). Although the earlier assumptions of a closed population and homogeneous mixing also hold in this case, the complexity of the dynamical process increases due to the addition of one more state. The agents, instead of only switching between susceptible and infected (as in SI model), tend to recover in the SIR epidemic model. The dynamics of the SIR model is controlled by two parameters: the infection rate, $\beta$, and the recovery rate, $\gamma$. The SIR model can be mathematically represented by the set of the following differential equations:
\begin{equation}
S + I + R =1
\end{equation}

\begin{figure}[t]
\begin{center}
\includegraphics[width=0.75\textwidth]{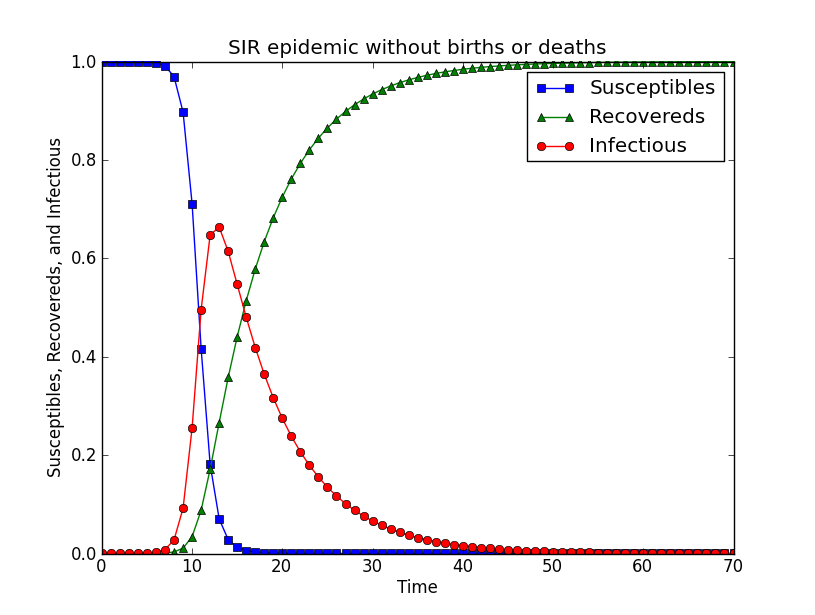}
\end{center}
\caption{Figure shows a typical SIR curve with $\beta=1.4$ and $\gamma=0.15$. The $y-$axis represents the population fraction of $S$, $I$ and $R$ species respectively and $x-$axis represents simulation time.}
\end{figure}

The population of susceptible nodes decreases in proportion to the number of encounters multiplied by the probability that each encounter results in an infection. The negative sign denotes that the population of S is decreasing. Similarly, we can describe the evolution of the other two states, I and R. Nodes become infected at a rate proportional to the number of encounters, and the probability of infection controlled by the parameter, $\beta$. Nodes recover at a rate proportional to the number of infected individuals, and the probability of recovery controlled by the parameter, $\gamma$:
\begin{equation}
\frac{dS}{dt}=-\beta SI,\quad \frac{dI}{dt}=\beta SI - \gamma I\quad and\quad \frac{dR}{dt}=\gamma I
\end{equation}
It would be interesting to analyze the spread of infection with respect to the susceptible individuals when there is a constant recovery (from eqn. 20). We calculate the variation in I with respect to S:
\begin{equation}
\frac{dI}{dS}=\frac{\gamma}{\beta}\frac{1}{S}-1
\end{equation}
The solution to the above equation with the initial conditions at $t=0$, $I \sim 0$ (negligible as compared to the population) and $S = 1$, is given by:
\begin{equation}
I=\int\Big (\frac{\gamma}{\beta}\frac{1}{S}-1\Big )dS = \frac{\gamma}{\beta}ln(S) - s +1
\end{equation}
In order to understand the rate of spread of infection in the population, we look at the rate equation for I from eqn. 20:
\begin{equation}
\frac{dI}{dt} = \beta SI - \gamma I = I(\beta S - \gamma)
\end{equation}
The above equation implies that the infection spreads if and only if $(\beta S - \gamma) > 0$. The epidemic dies out (the number of infected individuals decreases) if the above quantity is less than zero. Bifurcation occurs at the stationary state, when $\frac{dI}{dt} = 0$, which separates the above two regimes and corresponds to the epidemic threshold. For a network with average node degree $\bar{k}$, the rate of change of the susceptible population is given by:
\begin{equation}
\frac{dS}{dt} = -\beta\bar{k}SI
\end{equation}
The rate of change of the infected and simultaneously recovered individuals is similarly given by:
\begin{equation}
\frac{dI}{dt} = \beta\bar{k}SI - \gamma I \quad and \quad\frac{dR}{dt} = \gamma I
\end{equation}
Note that the rate of change of the recovered individuals remains the same as before. Substituting the value of I from eqn. 25 into eqn. 24, and solving for S gives us the following expression for S in terms of R,
\begin{equation}
S = -\int(\beta\bar{k}SI) dt = \exp(-\beta\bar{k}R)
\end{equation}
at time $t\rightarrow\infty$, $I \rightarrow 0$ and $S + R \rightarrow 1$. Population of recovered individuals, $R_{\infty}$ is given as $R_{\infty} = 1 - \exp(-\beta \bar{k} R_{\infty})$. Recovery of the individuals occur in the network if and only if the slope of $R_{\infty}\geq 1$ or $\beta \geq \bar{k}^{-1}$.

\section{Results: statistical and numerical analysis}
In this section, we present our main results on the studies of BTNs in India. First, we discuss the motivation of our study, and then we describe the datasets used for analysis. Before we discuss the primary results of this work in detail, we outline the network representation techniques specific to transportation terminology. 

\subsection{Motivation of the study}
We also saw earlier that any physical, chemical, biological or social system can be visualized as a complex network with constituting elements known as nodes, and the interactions between them identified as links. Based on the nature of the links, these networks can be broadly classified into virtual and spatial networks. In the former category, the links are physically absent, e.g., social networks or collaboration networks, whereas in the later case, the links are physically present, e.g., geographically embedded road or railway networks. In between these two broad classes, there exist networks in which the links, although physically absent, are still geographically constrained. The structure of the real-world networks such as bus or electric power grid are dependent upon the structure of the physically constrained, geographically embedded networks on which they grow and evolve. Therefore, our study would be incomplete if we do not explain the role constraints play in geographical embeddedness. We analyze this aspect by calculating the `dimensions' of these networks and check for self-similar patterns in them. 

In this study, we present the statistical analysis of the bus networks of six major Indian cities as graphs in $L$- and $P$-spaces, using concepts from network science. Although public transport networks such as airline and railway networks have been extensively studied, a comprehensive study on the structure and growth of bus networks is still lacking. In India, where BTNs play an important role in day-to-day commutation, it is of significant interest to analyze their topological structures and answer some of the basic questions on their growth, evolution, robustness and resiliency. Therefore, we do a comparative study of the bus networks of some of the major Indian cities, namely Ahmedabad (ABN), Chennai (CBN), Delhi (DBN), Hyderabad (HBN), Kolkata (KBN) and Mumbai (MBN). In order to understand the structure of these networks, we calculate various metrics, such as clustering coefficients, characteristic path lengths, degree-distribution and assortativity. We also simulate network robustness and resiliency by first removing nodes at random, followed by targeted removal based on degree, closeness and betweenness. Simulating node removals and simultaneously capturing the variation in their characteristic path lengths helps us in understanding nodal redundancy in these networks. Few studies on BTNs have although looked into the structural aspects in detail, the ESW network that we described earlier, is the only model that has looked into the aspect of network redundancy albeit due to geographical placements of the nodes. 

\subsection{Description of the dataset}
For this study, we use the bus routes as network datasets by considering bus stops as nodes and bus routes as links. The route details were obtained from the government websites of AMTS (ABN), MTC (CBN), DTC (DBN), APSRTC (HBN), CSTC (KBN), BEST (MBN) and Ahmedabad BRTS (Bus Rapid Transit System). It can be seen from Table 1 that the network sizes of all the cities are comparable to each other, except that of KBN, because CSTC is localized and operates as a subdivision of the West Bengal Surface Transport Corporation (WBSTC) that operates buses in the entire state. For computational and visualization purposes we parse the datasets as edge-lists, where the two adjacent columns are labeled as `source' and `target' respectively. In $L$-space representation, the values in the respective columns represent neighboring bus stops in a given route, whereas in $P$-space representation, the adjacent columns represent all possible transfers in a route for a fixed value in the column of `source' (see Figure 9).

\begin{figure}[hb!]
\begin{center}
\includegraphics[width=1\textwidth]{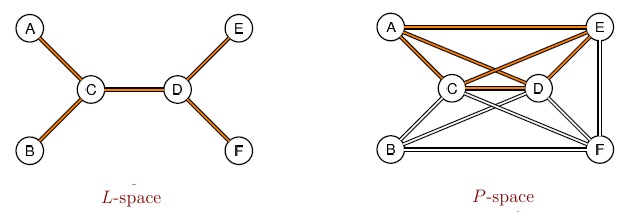}
\end{center}
\caption{Figure shows the $L$-space and $P$-space representation for transportation networks. If $A\rightarrow C\rightarrow D\rightarrow E$ represents a route in $L$-space, then in $P$-space representation we have all the possible transfers for the route, \emph{i.e.}, $A\rightarrow C\rightarrow D\rightarrow E, A\rightarrow D, A\rightarrow E, C\rightarrow E$.}
\end{figure}

\begin{sidewaysfigure}
\begin{center}
\includegraphics[width=1\textwidth]{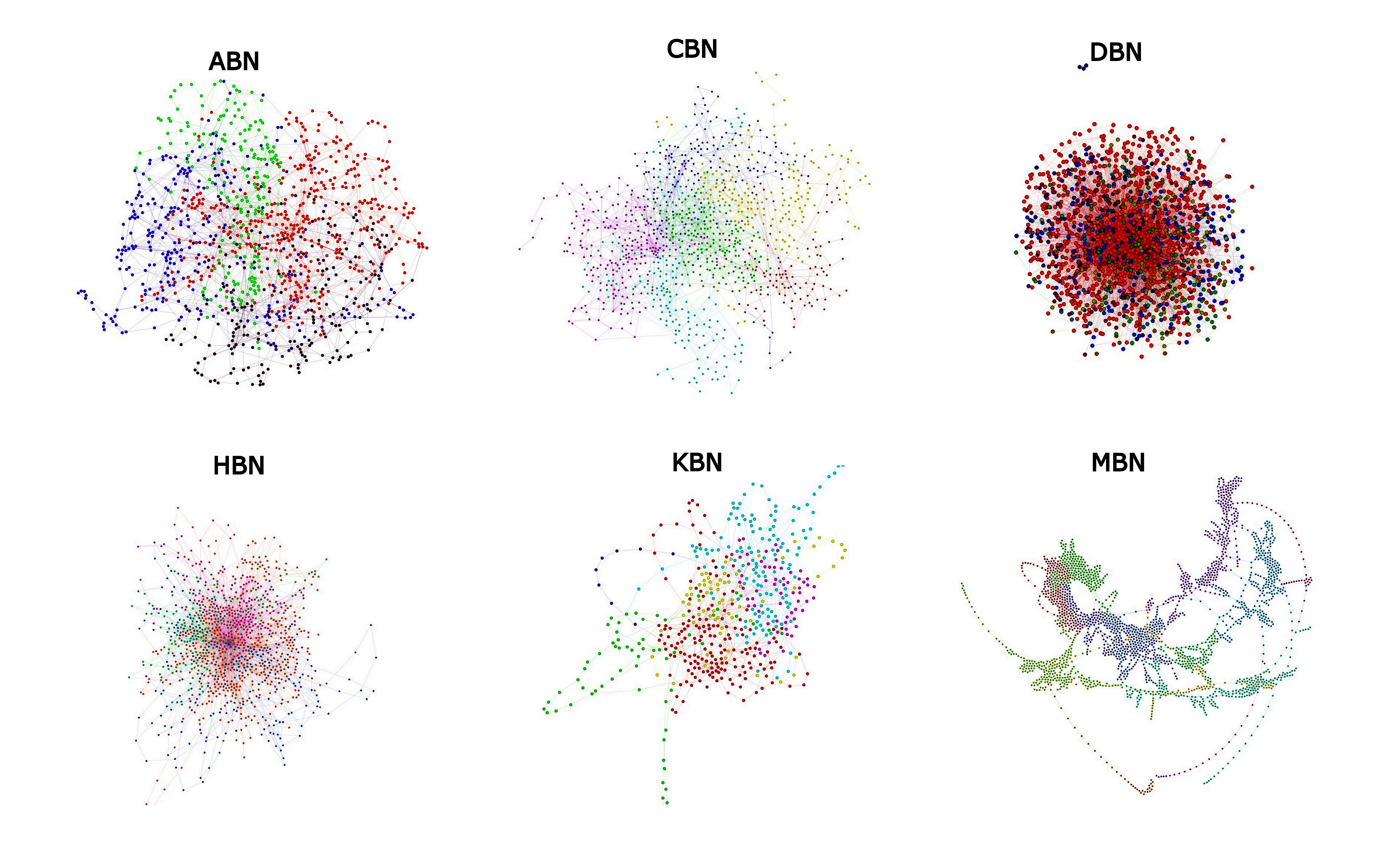}
\end{center}
\caption{Figure shows the network structure of the different bus routes in $L$-space, where each node represents a bus stop. The plots are generated using force directed algorithms, and the colours of nodes partition the networks into different communities. Where ABN, DBN and HBN show typical scale-free structure, observe the long routes present in CBN and MBN.}
\end{sidewaysfigure}

\subsection{Network topology}
Before we analyze the statistical properties of the different BTNs, we need to understand the relationship between the various network representation forms and their respective advantages. The most common is the $L$-space representation, where each bus stop is a node and a link between the nodes indicates that there is at least one route that services these two corresponding nodes consecutively (see Figure 9). In this representation, no multiple links are allowed between a pair of nodes, or $a_{ij}=1$. In cases of route overlaps, the element $a_{ij}$ is multiplied with a weight element $w_{ij}$, thus denoting strength of that link (or node): $s_{ij}=a_{ij}\times w_{ij}$. A different network representation is that of a bipartite graph that has been found to be useful in the analysis of cooperation networks. In this representation, also called $B$-space, both routes and stops are represented by nodes and each route node is linked to all bus stops that it services, and no direct links between nodes of same type occur. The neighbors of a given route node in this representation are all those stops that it services, while the neighbors of a given bus stop are all the routes that service it~\cite{guimera2004modeling, newman2003structure}. 

There are two one-mode projections of the bipartite graph of $B$-space. The projection to the set of station nodes is the $P$-space graph (see Figure 9), and the complementary projection to route nodes leads to the $C$-space graph. The $P$-space network representation has proven to be particularly useful in the analysis of PTNs~\cite{yang2014study, guimera2004modeling}. The nodes of this graph are bus stops, and they are linked if they are serviced by at least one common route. In this way, the neighbors of a $P$-space node are all those stops that can be reached without changing means of transport.

In order to get the essence of different network representations and their significance, we try to visualize the metric: characteristic length ($\textbf{l}_{ij}$), which in an $L$-space graph is the number of `hops' one has to make to travel between any two randomly chosen bus stops. When the network is represented in $P$-space, $\textbf{l}_{ij}$ signifies the number of bus changes one has to make in order to travel between any two randomly chosen bus stops. From a transportation perspective, less number of bus changes will imply a small-world property. Therefore, our calculation of the network metrics will strongly depend upon the network representation that is chosen

\subsubsection{$L$-space formulation}
\begin{table}[hb!]
\begin{center}
\begin{tabular}{|cccccccccc|}
\hline
{\bf Bus routes} & {\bf Nodes} & {\bf Edges} & {\bf $\textbf{l}_{ij}$} & {\bf $C_{av}$} & {\bf $\gamma$} & {\bf $\lambda$} & {\bf $r$} & {\bf $\bar{k}$} & {\bf $m$} \\ \hline
{\bf ABN}        & 1103        & 2582        & 5.59           & 0.19                         & 2.47           & -               & 0.07              & 3.67          & 4  \\
{\bf CBN}        & 1009        & 1610        & 8.73           & 0.07                         & 3.81           & -               & 0.12              & 24.58         & 6 \\
{\bf DBN}        & 1557        & 4287        & 5.51           & 0.18                         & 7.03           & 0.06      & 0.07              & 9.88           & 3 \\
{\bf HBN}        & 1088        & 2954        & 3.87           & 0.26                         & 3.52           & -               & -0.03             & 23.88      & 3 \\
{\bf KBN}        & 518         & 884         & 5.72           & 0.08                         & 4.96           & -               & -0.01             & 6.72          & 4 \\
{\bf MBN}        & 2267        & 3042        & 25.69          & 0.15                         & 8.53           & 0.13       & 0.18              & 10.38          &6\\ 
{\bf BRTS}      &129          &146		&9.55		&0.02			&*			&*		&0.26		&4.8		& * \\ \hline			
\end{tabular}
\caption{Tabular representation of the statistical data for the Ahmedabad BRTS and bus routes of six major Indian cities as graphs in $L$-space. $m$ represents the number of communities in these networks. For Ahmedabad BRTS, the dataset is very limited, which disables us from concluding its exact topological structure.}
\end{center}
\end{table}

In order to understand the topological structure of the various BTNs in India, we formulate these networks as graphs in $L$-space~\cite{chatterjee2014scaling, chatterjee2015statistical}. We calculate the various network metrics and tabulate our results in Table 1. We also use force-directed algorithms to visualize the topological structure of these networks (see Figure 10; for Ahmedabad BRTS, refer Figure 11). The figure compares the structural construct of the networks. We can clearly observe the nature of connectivity between the nodes in the different networks. While DBN is densely packed, CBN, HBN and KBN are sparse. The network structure of MBN is particularly striking. The long branches with multiple intermediate nodes, as seen in Figure 10, cause the characteristic path-length, $\textbf{l}_{ij}$ of MBN to increase abnormally (see Table 1). We also calculate the modularity of the networks to identify community structure (Girvan-Newman algorithm)~\cite{girvan2002community}. Networks with high modularity have dense connections between the nodes within the same modularity class, but weak connections between nodes in different modularity class. In order to identify communities, we colour-code the nodes based upon the modularity classes. Community detection in bus networks helps us in identifying the different zones of operation. As large as six communities were identified for CBN and MBN, whereas fewer (four or less) communities were identified for ABN, DBN, HBN and KBN. 

\subsubsection{$P$-space formulation}
An interesting result is seen in the characteristic path length values of the different networks in Table 1. The magnitudes of $\textbf{l}_{ij}$ for CBN, MBN and Ahmedabad BRTS show peculiarly high values in $L$-space. Therefore, these values do not portray the complete picture of these networks. We carry out a $P$-space statistical analysis for CBN, MBN and Ahmedabad BRTS, and tabulate our results in Table 2. Since, we connect every stop to every other stop along a route, notice the large increment in the number of links for these networks in $P$-space representation. The $P$-space analysis for CBN, MBN and Ahmedabad BRTS reveals small-world phenomena due to very low characteristic path length and very high clustering. 

We also plot the topological structure of the networks in Figure 11. By employing clustering algorithm, we can clearly observe strong knit clusters in the network. Observe the central core of the network as one large strongly connected component and the strong-knit communities of neighbouring nodes in the periphery. These clusters on the periphery tend to belong to one single route with hubs present in the central core. Therefore, they show strong connectivity within themselves, as well as between them and the central core.

Also notice from the the tables, (Table 1 and 2) the stark changes in the value of degree-assortativity coefficient. We can observe a positive correlation between $\textbf{l}_{ij}$ and $r$, meaning strongly assortative networks tend to have longer characteristic path lengths and vice-versa. 

\begin{table}[hb]
\begin{center}
\begin{tabular}{|ccccccccc|}
\hline
{\bf Bus routes} & {\bf Nodes} & {\bf Edges} & {\bf $l_{ij}$} & {\bf $C_{av}$} & {\bf $\gamma$} & {\bf $\lambda$} & {\bf $r$} & {\bf $\bar{k}$} \\ \hline
{\bf CBN}        & 1009        & 21858        & 2.42           & 0.74                         & 3.81           & -               & -0.12              & 43.08          \\
{\bf MBN}        & 2267        & 110941        & 2.51          & 0.68                         & 5.28           & -       & 0.09              & 107          \\ 
{\bf BRTS}      &129          &3193		& 1.71		&0.86			&*			&*		& -0.04		&49.33		\\ \hline			
\end{tabular}
\caption{Tabular representation of the statistical data for CBN, MBN and Ahmedabad BRTS as graphs in $P$-space. Note the significant difference in the statistical properties of these networks as compared to their counterparts in $L$-space.}
\end{center}
\end{table}

\begin{figure}[t]
\begin{center}
\includegraphics[width=0.85\textwidth]{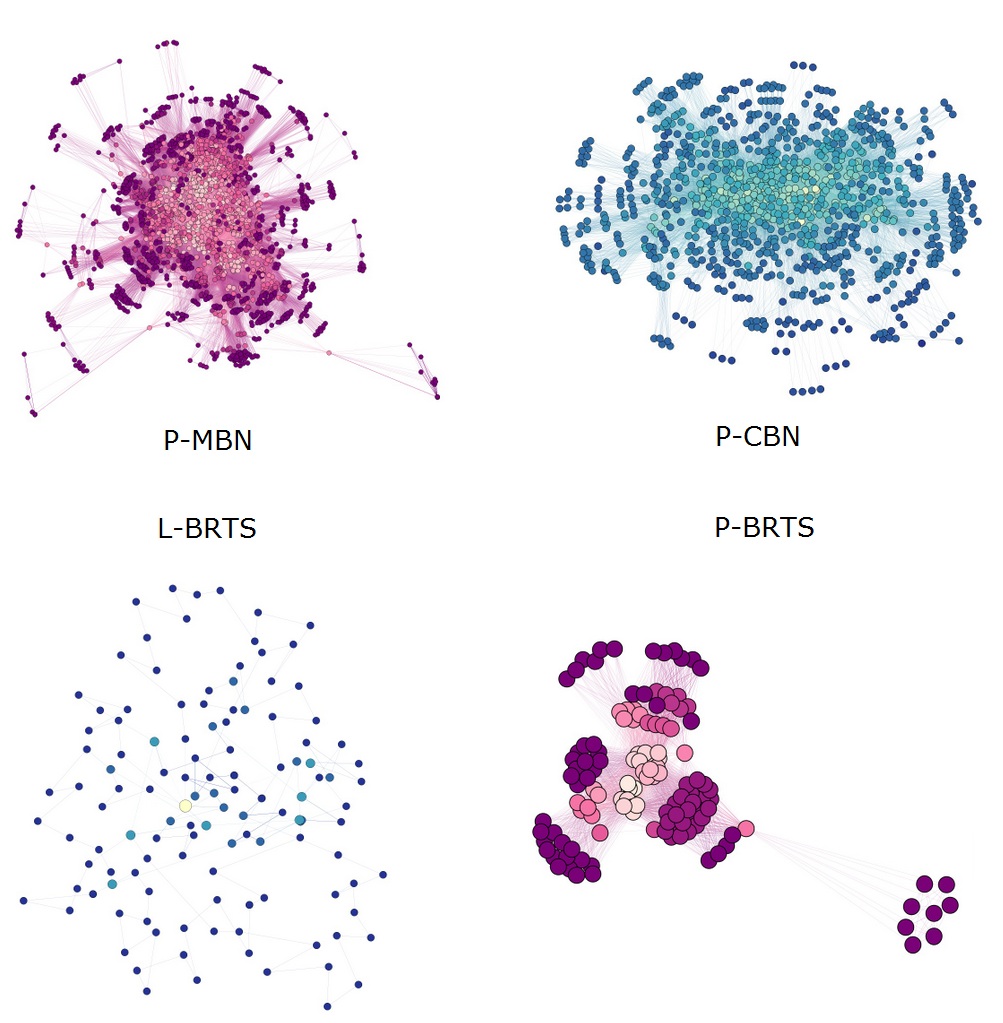}
\end{center}
\caption{Figure shows the topological structure of the networks: CBN, MBN and Ahmedabad BRTS in $P$-space. We also plot the network structure of Ahmedabad BRTS in $L$-space. Notice the strong-knit peripheral communities and the central core.}
\end{figure}

\subsection{Local network properties}
In this section, we present our results on the statistical analysis of the BTNs. First, we look into the local network properties: node-degree distribution and node clustering coefficients, then we will try to visualize the big picture from the entire network's perspective. Some of the interesting network characteristics that we will look at are: network centralities (closeness and betweenness), characteristic path length distribution, redundancy in the topological structure of the networks and the way hubs are connected in the network, $i.e.$, degree-assortativity coefficient. 

\subsubsection{Network node-degree distribution}
In Figure 12, we plot the weighted degree-distribution (or strength-distribution) for all the networks in $L$-space representation on a double logarithmic scale (for Ahmedabad BRTS, see Figure 13). The degree-distribution patterns show heavy-tailed characteristics, as can be seen from the plots. More specifically, ABN, CBN, HBN and KBN show power-law behaviour, whereas DBN and MBN have an exponential distribution. None of these graphs follow pure power-law distribution. Therefore, the degree-distribution function can be represented as a truncated power-law, $P(s)\sim\exp(-s/s_{min})k^{-\gamma}$ for ABN, CBN, HBN and KBN ($s_{min}\Rightarrow$ degree-strength cut-offs), whereas for DBN and MBN, the degree-distribution function follows an exponential decay, $P(s)\sim\exp(-\lambda s)$, where $\gamma$ and $\lambda$ are respectively the decay exponents (see Table 1). 

\begin{figure}[hb]
\begin{center}
\includegraphics[width=1\textwidth]{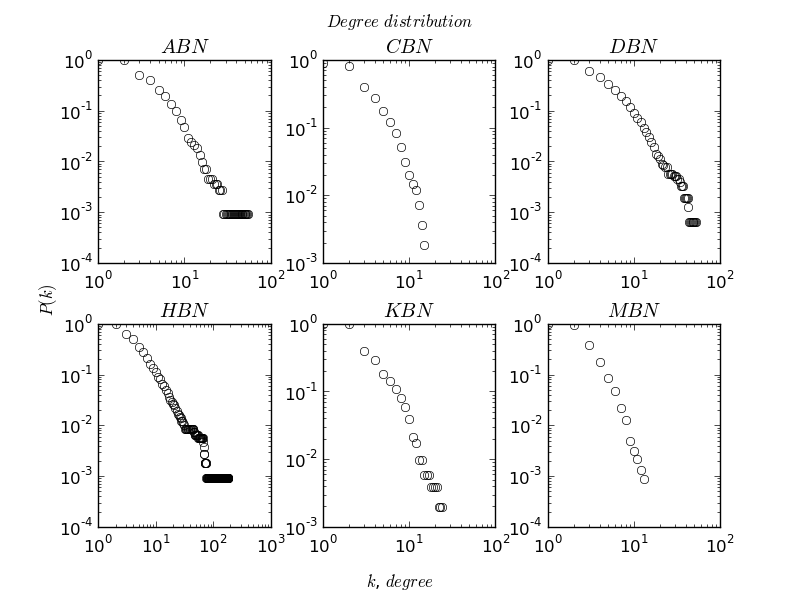}
\end{center}
\caption{Figure shows the weighted degree-distribution, $P(s)$ on a double logarithmic scale for the networks in Table 1 (except Ahmedabad BRTS).}
\end{figure}

As we discussed earlier, the higher organization of real world networks usually leads to slower decaying distributions. Typical classes of such networks have either exponential or power law tails. Both exponential and power law decay of the degree-distribution can be modeled by assuming non-equilibrium growth processes of the networks~\cite{barabasi1999emergence, deng2011exponential}. An interesting observation from the plots for CBN and MBN in $L$- and $P$-spaces is that in $P$-space both the networks tend to show slower decay (compare density functions for MBN in both $L$- and $P$-spaces). The degree-exponent for CBN in both $L$- and $P$-space remains the same, however in $P$-space, CBN shows a sharper cut-off. Also observe that the degree-distribution pattern for Ahmedabad BRTS shows a discrete function in $L$-space, imitating a heavy-tail. Whereas in $P$-space, it decays rapidly, imitating either an exponential, or most-likely a log-normal density function. 

The equation for the power-law or exponential fits in Tables 1 and 2, and Figures 12 and 13 are calculated using Maximum Likelihood Estimation (MLE), and the Kolmogorov-Smirnov test is employed to check for goodness of fit \cite{clauset2009power}.

\begin{figure}[hb!]
\begin{center}
\includegraphics[width=1\textwidth]{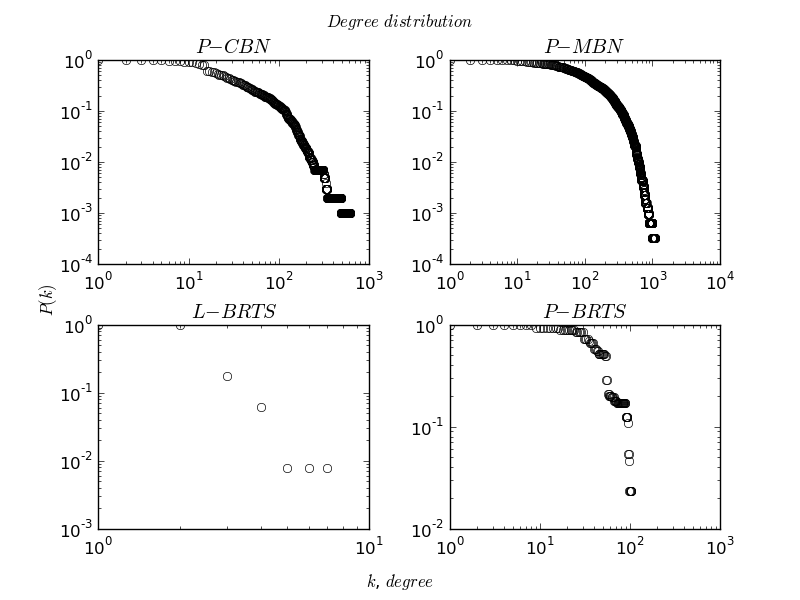}
\end{center}
\caption{Figure shows the degree-distribution, $P(k)$ on a double logarithmic scale. The top panel shows the degree-distributions for CBN and MBN in $P$-space, and the lower panel for Ahmedabad BRTS in $L$- and $P$-spaces.}
\end{figure}

The other important results are the degree-strength distribution patterns. In order to check for correlations between node degree, $k$ and node weighted-degree, $s$, we plot them on a double-logarithmic scale (see Figure 14). Interestingly, ABN shows a strong power-law correlation, as $s\sim  k^{\beta}$ with $\beta = 1.27$ and $R^2 = 0.91$, whereas the other networks fail to show such strong relationships (CBN, KBN and HBN show similar relationships with $\beta \sim 1.44 - 2.08$, however with very low correlation coefficient, $R^2 \sim 0.58 - 0.76$). The degree-strength distribution for ABN and HBN shows a linear form, implying high degree nodes are also the ones that handle handle heavy traffic. The degree-distribution in case of ABN and HBN have the power-law exponent, $\gamma$ as $2.47$ and $3.52$ respectively, whereas the degree-strength exponent, $\beta$ is found to be $1.27$ and $1.09$. This implies that the strength of a node increases faster as compared to its degree, which basically shows a sense of order in both ABN and HBN, where higher degree nodes, for example, large or important bus stops, handle heavy traffic as majority of the routes pass through them. This is definitely missing in the other networks, where the edge weights or routes seem to be more randomly distributed. Also observe in case of HBN, the almost linear correlation between degree and strength results in a star-like topology in the $L$-space network representation (see Figure 10).

\begin{figure}[hb!]
\begin{center}
\includegraphics[width=1\textwidth]{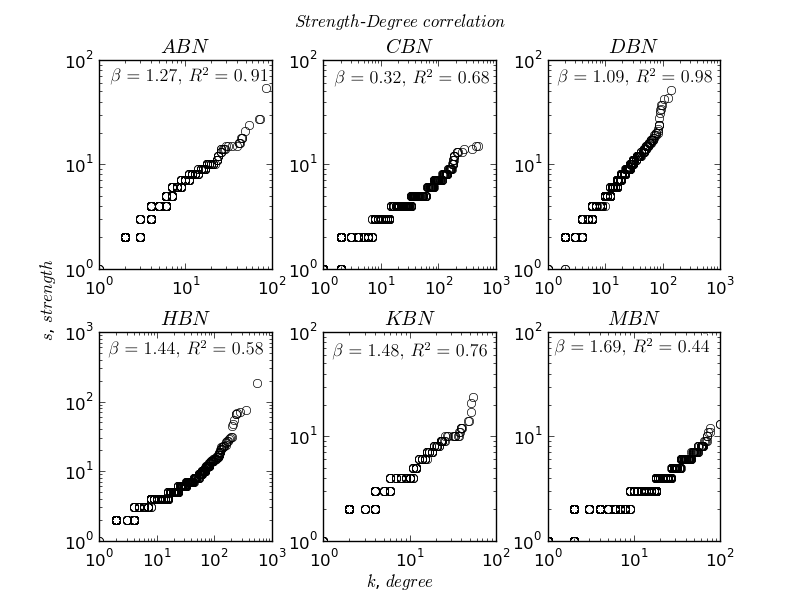}
\end{center}
\caption{Figure shows the degree-strength distribution on a double logarithmic scale for the six networks in $L$-space. The relationship between strength and degree follows a power-law pattern, with $s\sim k^{\beta}$. We outline the various values for the exponent $\beta$ with the corresponding goodness of fit $R^2$ in the plots.}
\end{figure}

\subsubsection{Node clustering coefficient}
In Tables 1 and 2, we present the values of the mean clustering coefficient in $L$- and $P$-spaces. The highest absolute values of the clustering coefficient are found in $P$-space, where their range is given by $C = (0.7, 0.8)$, whereas in $L$-space, $C = (0.01, 0.26)$. This is not surprising, since in $P$-space, each route gives rise to a fully connected (complete) subgraph between all of its stations. In $L$-space the correlation between the clustering coefficient and node-degree is found mostly to be weak, except for ABN and HBN, where $C(k)\sim k^{-0.75},\quad R^2 = 0.78$ and $C(k)\sim k^{-0.59},\quad R^2 = 0.83$ respectively. DBN shows weak correlation, as $C(k)\sim k^{-0.82},\quad R^2 = 0.63$, whereas for CBN, KBN and MBN, no conclusive relationships are found. However, in $P$-space the clustering coefficient of a node is strongly correlated with the node degree. In Figure 15, we plot the variation in clustering coefficient with node degree for CBN in $P$-space. We can clearly see from the plot that the best fit line nicely obeys a power-law relationship. 

The variation of clustering in nodes with the inverse power of degree shows evidence of hierarchical structure in the network. Networks that have high degrees are the hub nodes and they tend to form large number of small and long connections. This reduces the probability of these nodes to form local clusters, whereas nodes with low degrees are more likely to form clusters. Network hierarchy emerges when clusters of low degree nodes tend to connect to high degree hubs. Thus, hubs act as central coordinating agents to which clusters of lower rank agents (low degree nodes) directly interact with, while also interacting among themselves. 

\begin{figure}[hb!]
\begin{center}
\includegraphics[width=0.5\textwidth]{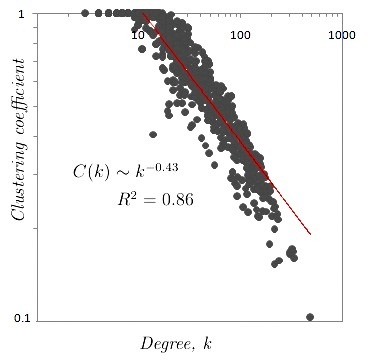}
\end{center}
\caption{Figure shows the variation in clustering coefficient with the node degree in $P$-space representation of CBN.}
\end{figure}

\subsection{Global network properties}
In this section, we discuss about those network metrics that largely depend upon the topology of the network. Properties like node centrality or characteristic path length depend upon the network size and the connectivity among the nodes. These metrics provide us with global characteristic features of the network. 

\subsubsection{Characteristic path length and network redundancy}
As was mentioned earlier, the characteristic path length $\textbf{l}_{ij}$ is the average length of the shortest path between any pair of random nodes, $n_i$ and $n_j$ in a graph. The metric, $\textbf{l}_{ij}$ is well-defined if and only if the nodes $n_i$ and $n_j$ belong to the same connected component of the graph. In general, the characteristic path length distributions obtained in $L$- and $P$-spaces are nicely described by an asymmetric unimodal distribution~\cite{newman2003structure}:
\begin{equation}
P(\textbf{l}_{ij}) = A\textbf{l}_{ij}\exp(-B{\textbf{l}_{ij}}^2+C\textbf{l}_{ij})
\end{equation}
where A, B, and C are the parameters.

For all our datasets, we found a similar distribution of the characteristic path lengths, with varying magnitudes of the parameters A, B and C. We plot the characteristic path length distribution function in Figure 16. 

\begin{figure}[t]
\begin{center}
\includegraphics[width=0.5\textwidth]{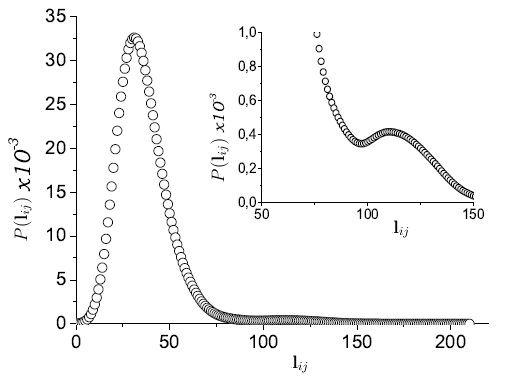}
\end{center}
\caption{Figure shows the characteristic path length distribution in $L$- and $P$-space for the BTNs.}
\end{figure}

The peak of the plot in Figure 16 tells us the magnitude of the characteristic path length for the entire network. This magnitude interestingly turns out to be the inverse of closeness centrality ($C_C$). Interesting results are obtained when we compare the variation of the characteristic path length with that of the network size by simulating random as well as targeted node removals. In Figure 17, we plot the response of the network's characteristic path length, $\textbf{l}_{ij}$ to random and systematic perturbation. We simulate the robustness and resiliency of the networks by modeling perturbations as node removals. Due to their strong assortative nature, MBN and CBN disintegrate into separate entities very quickly, whereas the other networks remain connected upto atleast $4\%$ of node removals. It is observed that in all the cases the targeted node removals are crucial for the network to remain connected. In the regime of $p_i \leq 4\%$, a closer look reveals that the magnitude of $\textbf{l}_{ij}$ does not change much (at most it increases by one `hop'). 

This leaves before us an important question: Are there redundant nodes in the network? When we analyzed the BTNs in $L$- and $P$-spaces (Tables 1 and 2), we observed that in cases where $\textbf{l}_{ij}$ had particularly high values in $L$-space (\emph{i.e.}, large number of hops) it got reduced to very low values in $P$-space (\emph{i.e.}, very few transfers). This provides us with an option of optimizing the BTNs by removing the redundant nodes from the network. 

\begin{sidewaysfigure}
\begin{center}
\includegraphics[width=1\textwidth]{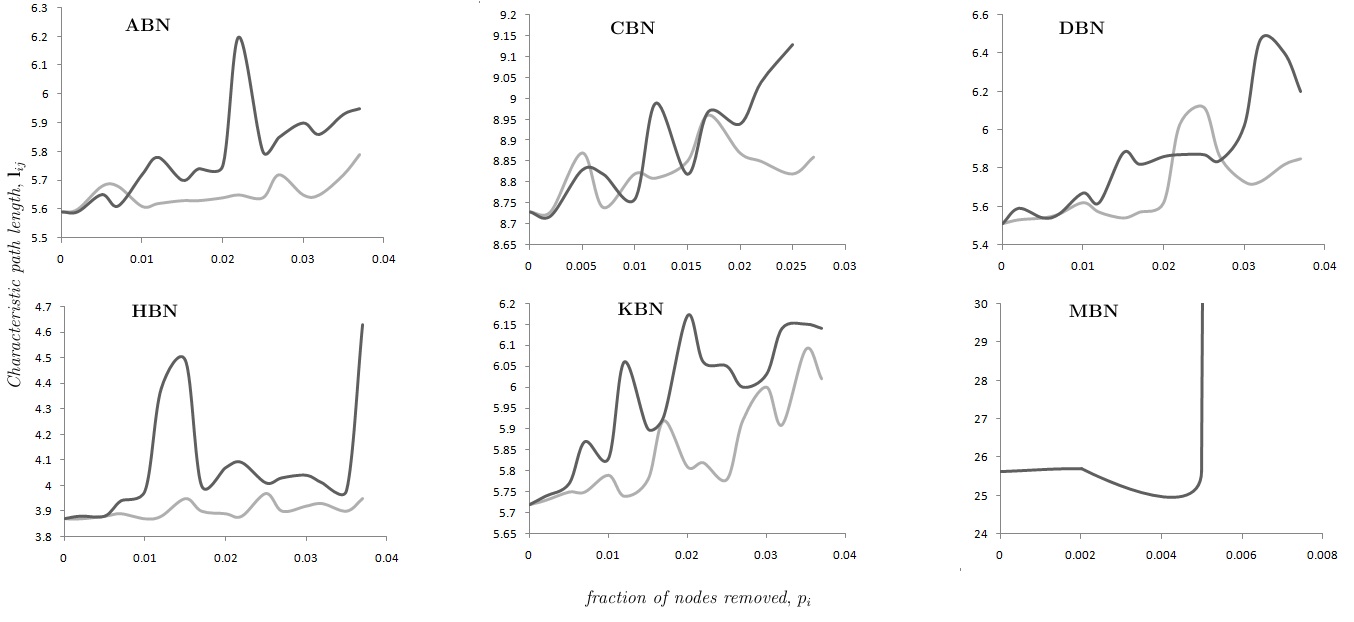}
\end{center}
\caption{Figure shows the variation in $\textbf{l}_{ij}$ with network size upon random and targeted node removals. The dark line represents degree-based node removals, and the light line represents random node removals. The X-axis represents the fraction of nodes removed.}
\end{sidewaysfigure}

\subsubsection{Network centralities}
The presence of degree-heterogeneity results in disparity between nodes in a network. Certain nodes in a network are more `central' than the others. In this context, we consider betweenness and closeness as our node-centrality because they play a crucial role from a transportation perspective. $C_C$ is a measure of a node's relative importance in the network due to the existence of shortest paths from that particular node to every other node in the entire network. $C_B$ on the other hand acts as a bridging node connecting different parts of the network together. When traveling from one node to the other, it is often beneficial to get to the node with the highest value of $C_C$ first if a direct path does not exist between the origin-destination pair. Thus, the centralities, betweenness and closeness tell us the relative importance of nodes in the network. Betweenness centrality of any node is calculated as:
\begin{equation}
C_B (i) = \Sigma_{s\neq i \neq t}\frac{\sigma_{s, t}(i)}{\sigma_{s, t}}
\end{equation}
where $\sigma_{s, t}$ is the number of shortest paths connecting $s$ to $t$ and $\sigma_{s ,t}(i)$, number of shortest paths connecting $s$ to $t$ but passing through $i$. Likewise, closeness centrality for any node is calculated as:
\begin{equation}
C_C (i) = \Sigma_i \frac{1}{a_{ij}}
\end{equation} 
The average closeness is the harmonic mean of the shortest paths from any node to every other node. In weighted networks, usually the edge weights are considered as cost functions. Therefore, larger the edge weight, lesser is the node's closeness, as the cost of travel would be large. However in our case, the edge weights play an altogether different role signifying the `ease' of travel, hence, we take the inverse of edge weights during the calculation of weighted $C_C$ as in collaboration networks, given by $C^{w}_C (i) = \min\Sigma_i (\frac{1}{w_{ij}})$.

\begin{figure}[hb!]
\begin{center}
\includegraphics[width=1\textwidth]{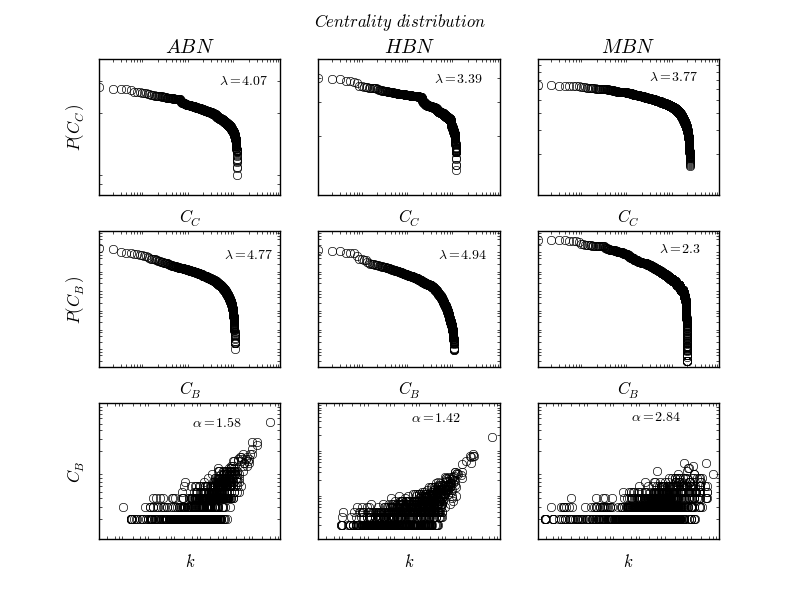}
\end{center}
\caption{Figure shows centrality distribution for betweenness ($C_B$) and closeness centralities ($C_C$), with the decay exponent $\lambda$ (inset) for ABN, HBN and MBN. The plots in the last row show degree-betweenness dependency with exponent $\alpha$ (inset).}
\end{figure}

In Figure 18, we plot the centrality distributions (closeness and betweenness), $P(C_C)$ and $P(C_B)$ in the first two rows for ABN, HBN and MBN on a double logarithmic scale. Only three specific bus networks are chosen since they capture the range of all network topologies shown by the six cities. While ABN shows more of a strict power-law (due to low exponential cut-off), HBN shows a truncated power-law, and MBN, an exponential form of degree-distribution. We find that the distribution function follows an exponential decay, given by $P(C_C)\sim\exp(-\lambda C_C)$ (similarly for $C_B$), where the value of the exponent $\lambda$ is shown in each of the plots. The centrality distribution plots decay rapidly, implying randomness in the distribution of central nodes in the network. Thus, new nodes do not \emph{preferentially} choose existing central nodes to connect to, rather they connect to their existing neighbors. In the last row, we plot the variation of betweenness centrality with the degree of a node which follows a power-law relationship, given as $C_B\sim k^\alpha$, with the magnitude of the exponent $\alpha$ also shown in the plots. A close observation reveals that nodes with high betweenness certainly have high degrees, however the reverse is not true. 

\subsubsection{Network assortativity}
The degree-distribution functions help us in understanding how different nodes tend to connect to each other in a network. However, they do not tell us how nodes with similar properties connect with themselves. In order to understand the connectivity patterns among groups of similar nodes, we calculate the degree-assortativity coefficients ($r$) for these networks (see Table 1 and 2). The degree-assortativity or the Pearson correlation coefficient of degree between pairs of connected nodes is given by:
\begin{equation}
\Sigma_{jk}\frac{jk(e_{jk}-q_j q_k)}{\sigma^{2}_q}
\end{equation}
where $e_{jk}$ is the joint probability distribution of the remaining degrees of the two vertices at either end of a randomly chosen edge, with $\Sigma_{j, k} e_{jk} =1$ and $\Sigma_j e_{jk} = q_k$. Here, $q_k$ is the normalized degree-distribution of the remaining degrees, and $\sigma^{2}_q$ is the variance of the distribution $q_k$, given by~\cite{newman2002assortative}:
\begin{equation}
\sigma^{2}_q = \Sigma_k k^2 q_k - [\Sigma_k kq_k]^2
\end{equation}

Since assortativity or assortative mixing is a preference for a network's nodes to attach to others that are similar in some way, we can visualize their topology by plotting the degree-correlation matrix. In Figure 19, we plot the degree-degree correlation matrix of the six BTNs in $L$-space, and for CBN and BRTS in $P$-space. Since CBN and MBN are strongly assortative in $L$-space, the patches in their plots are uniformly distributed, signifying all similar degree-type nodes connect to all other similar degree-type nodes in varying strength. Whereas for networks such as, ABN, DBN, KBN and HBN, which are either disassortative or weakly assortative we find high density patches in certain regions, and almost no patches in some regions as well. The high density patches represent close-knit groups between those similar nodes and absence of a patch indicates no link connectivity. The patterns are more interesting in the $P$-space topologies, where high density patches represent clusters because of large number of connections in between them (also see Figure 11).

\begin{sidewaysfigure}
\begin{center}
\includegraphics[width=0.9\textwidth]{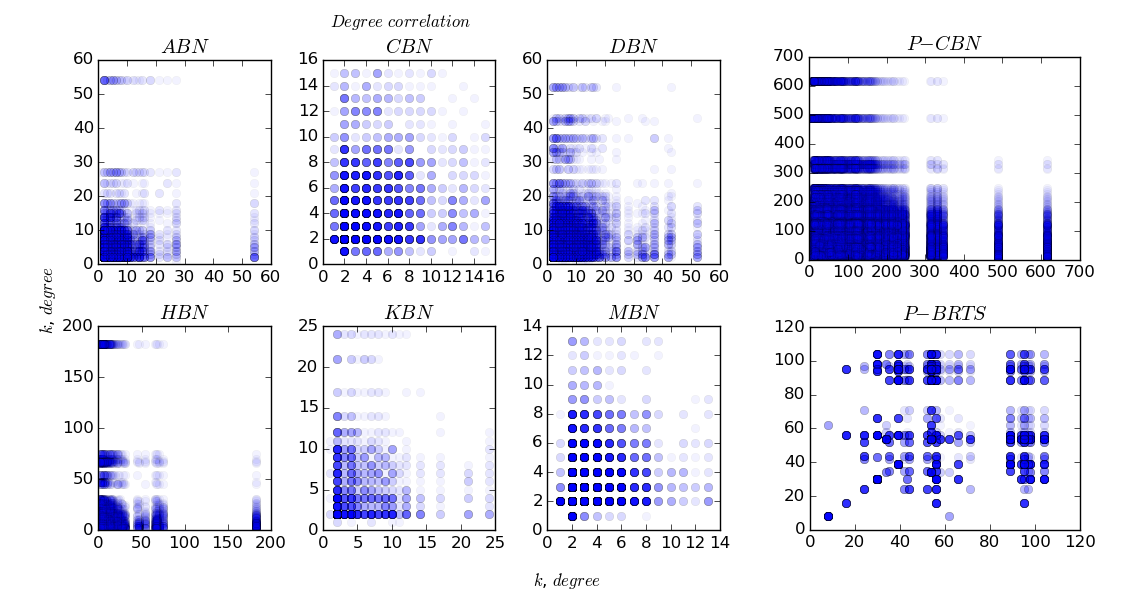}
\end{center}
\caption{Figure shows degree-degree correlation for the networks in $L$-space representation, and for Ahmedabad BRTS and CBN in $P$-space. The density of the patches represent the density of connections between `similar' nodes.}
\end{sidewaysfigure}

\subsection{Sub-network topology}
In the above sections, we presented our analysis on the complete datasets for the six bus routes. However, in cities certain bus routes tend to be special, like express service or night service, which neither travel along all the bus stops nor are they very large in number. In this section, we analyze those sub-networks for CBN and MBN and visualize their network characteristics. Their analysis in the current context solves the following two purposes: (\textbf{i}) evidence of self-similarity or fractal properties in the networks, and (\textbf{ii}) to understand how the absence of nodes changes the network topology. We define a sub-network (sub-graph) of $G=(N, L)$ as the graph $G_i = (N_i, L_i)$, such that $G_i\subset G$. The set of bus stops for the special routes will obviously be chosen from the exhaustive set of all the bus stops, therefore $N_i\subset N$. Since every route in the special case (service) can be identified as a derivative of the ordinary routes, $L_i\subset L$. 

\subsubsection{Network self-similarity}
A self-similar object is exactly or approximately similar to a part of itself. Many objects in the real world, such as coastlines or snowflakes, are statistically self-similar: parts of them show the same statistical properties at many scales~\cite{mandelbrot1967long}. In Figure 20, we plot the degree-distribution for the sub-networks of CBN and MBN in $L$-space. $G_1$ represents the night service network and $G_2$, the express service for both CBN and MBN. We observe that the plots show similar patterns for the density function as the original networks (see Figure 12). This motivates us to look for self-similarity in these networks. 
 
\begin{figure}[hb!]
\begin{center}
\includegraphics[width=1\textwidth]{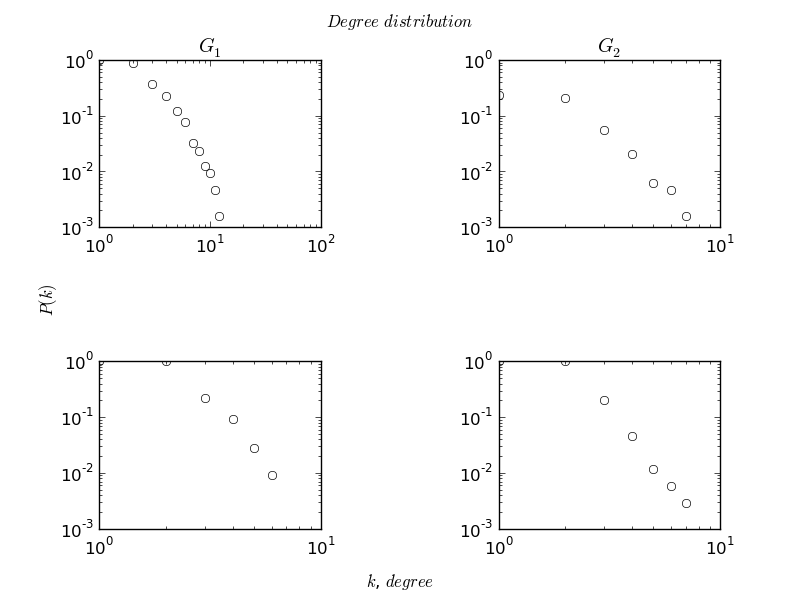}
\end{center}
\caption{Figure shows the degree-distribution plots for the sub-networks, $G_1$ and $G_2$ on a double-logarithmic scale.}
\end{figure}

\subsubsection{Fractal dimensions}
Analysis of a variety of real complex networks shows that they are self-similar on all length scales. Thus, our earlier observations on self-similarity were not trivial. The concept of self-similarity is ingrained in the complex network topologies. The small-world property can be mathematically expressed by the slow increase in the characteristic path length of the network, with the network size (WS model):
\begin{equation}
\textbf{l}_{ij}\sim\ln(N)\Rightarrow N\sim\exp(\textbf{l}_{ij} /\textbf{l}_0)
\end{equation}
where $\textbf{l}_0$ is a characteristic length.

For a self-similar structure, a power-law relation is expected rather than the exponential relation above. From this fact, it would seem that the small-world networks are not self-similar under length-scale transformations~\cite{song2005self}. To investigate self-similarity in complex networks, we use the box-counting algorithm and renormalization technique. For each side of length $l$, boxes are chosen randomly until the network is covered. A box consists of nodes separated by a distance $l_{ij} < l$. Each box is then replaced by a node. The renormalized nodes are connected if there is at least one link between the un-renormalized boxes. This procedure is repeated until the network collapses to one node (one box). Each of these boxes has an effective mass (the number of nodes in it) which can be used, to measure the fractal dimension of the network. We graphically express the box-counting algorithm in Figure 21.

\begin{figure}[hb!]
\begin{center}
\includegraphics[width=0.8\textwidth]{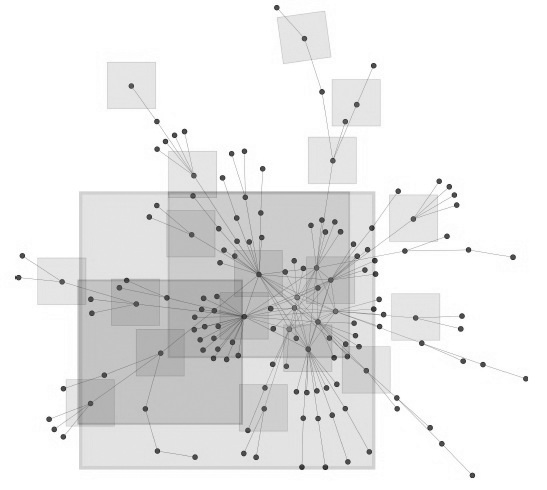}
\end{center}
\caption{Figure shows the box-counting algorithm to calculate fractal dimensions of complex networks.}
\end{figure}

We use the above algorithm to calculate the fractal dimensions of the BTNs. In Figure 22, we plot our results on a double-logarithmic scale, $N_{b} (l)$ against $l$. The above relation scales as $N_b (l)\sim l^{-\delta}$, where $\delta$ is the fractal dimension of the networks. 

Fractal dimensions were first applied as an index characterizing complicated geometric forms. For sets describing ordinary geometric shapes, the theoretical fractal dimension equals the set's familiar Euclidean or topological dimension. Thus, $\delta = 0\Rightarrow$ sets describing points (0-dimensional sets); $\delta = 1\Rightarrow$ sets describing lines (1-dimensional sets having length only); $\delta = 2\Rightarrow$ sets describing surfaces (2-dimensional sets having length and width); and $\delta =3\Rightarrow$ sets describing volumes (3-dimensional sets having length, width, and height). Our analysis shows a wide range of $\delta$, mostly within the limit $(2, 3)$. Interestingly, HBN has a fractal dimension, $\delta > 3$ and MBN, $\delta < 2$. We earlier claimed that bus networks are constrained by the geographical limitations. These networks although not having tangible links, need an underlying network to grow and evolve, which in this case is that of road networks (which are planar, with topological dimension = 2). Therefore, relating the fractal dimensions of these networks to the underlying topology of the road networks reveals the presence of constraints in them. As most networks have a fractal dimension, $2<\delta <3$, they signify a volumetric spanning (even though while being constrained in a 2-dimensional space) instead of a planar one. The structure of HBN as we had discussed earlier is a star type topology, thus allowing more structural freedom in the route connectivity as compared to others. Therefore, it has a fractal dimension, $\delta > 3$. MBN however, has a dimension $\delta<2$, signifying a more linear (stretched) evolution. 

\begin{figure}[t]
\begin{center}
\includegraphics[width=1\textwidth]{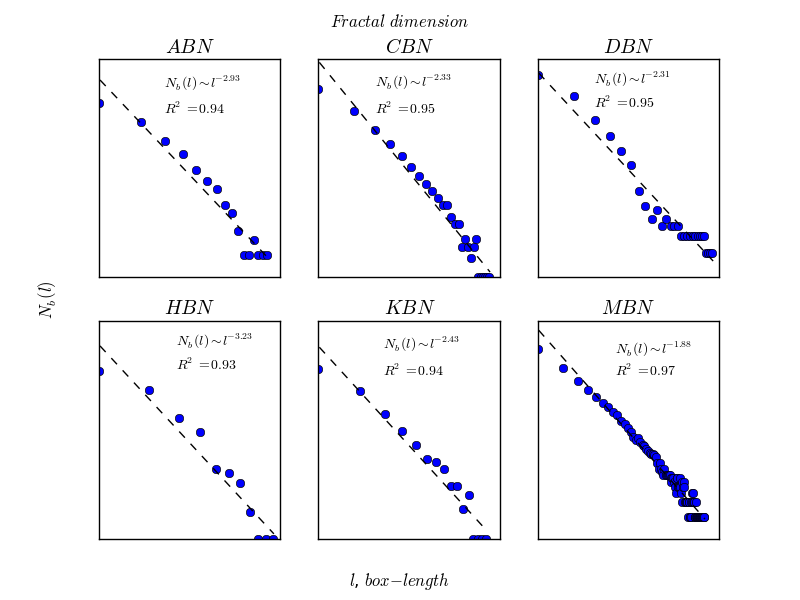}
\end{center}
\caption{Figure shows the fractal dimensions of the BTNs on a double-logarithmic scale. The fractal dimension, $\delta$ and the magnitudes of the corresponding goodness of fits $R^2$ are provided in the plots.}
\end{figure}

\subsection{Network dynamics: numerical results}
Until now all our analysis was focused on the static structural properties of the networks. In this section, we numerically simulate dynamical processes: SI and SIR to understand the role played by the network structure in general. 

\subsubsection{Network percolation: SI model}
In Figure 20, we simulate the diffusion dynamics in the networks. The plots show the Cumulative Distribution Function (CDF) of infection transmission in the network (Y-axis) with respect to simulation time (X-axis). As we saw in eqn. 18, the analytical solution for the SI model gives a logistic curve. However, the rate of diffusion or the slope of the curve and the saturation thresholds will be different for different networks due to their underlying topology. It is interesting to see how the various network metrics affect information diffusion in the following networks. We observe that the characteristic path-length $\textbf{l}_{ij}$ has a direct effect on the diffusion rate in these networks. The above observation is quite obvious, as the metric $\textbf{l}_{ij}$ tells us the number of hops that are required to navigate the entire network. From Figure 23, we can observe the simulation time for MBN and HBN by looking at the steepness of the plots. While HBN exhibits the steepest ascent, MBN takes the longest simulation time, which directly correlates to the magnitudes of the characteristic path lengths of MBN and HBN from Table 1. 

\begin{figure}[hb]
\center
\includegraphics[width=1\textwidth]{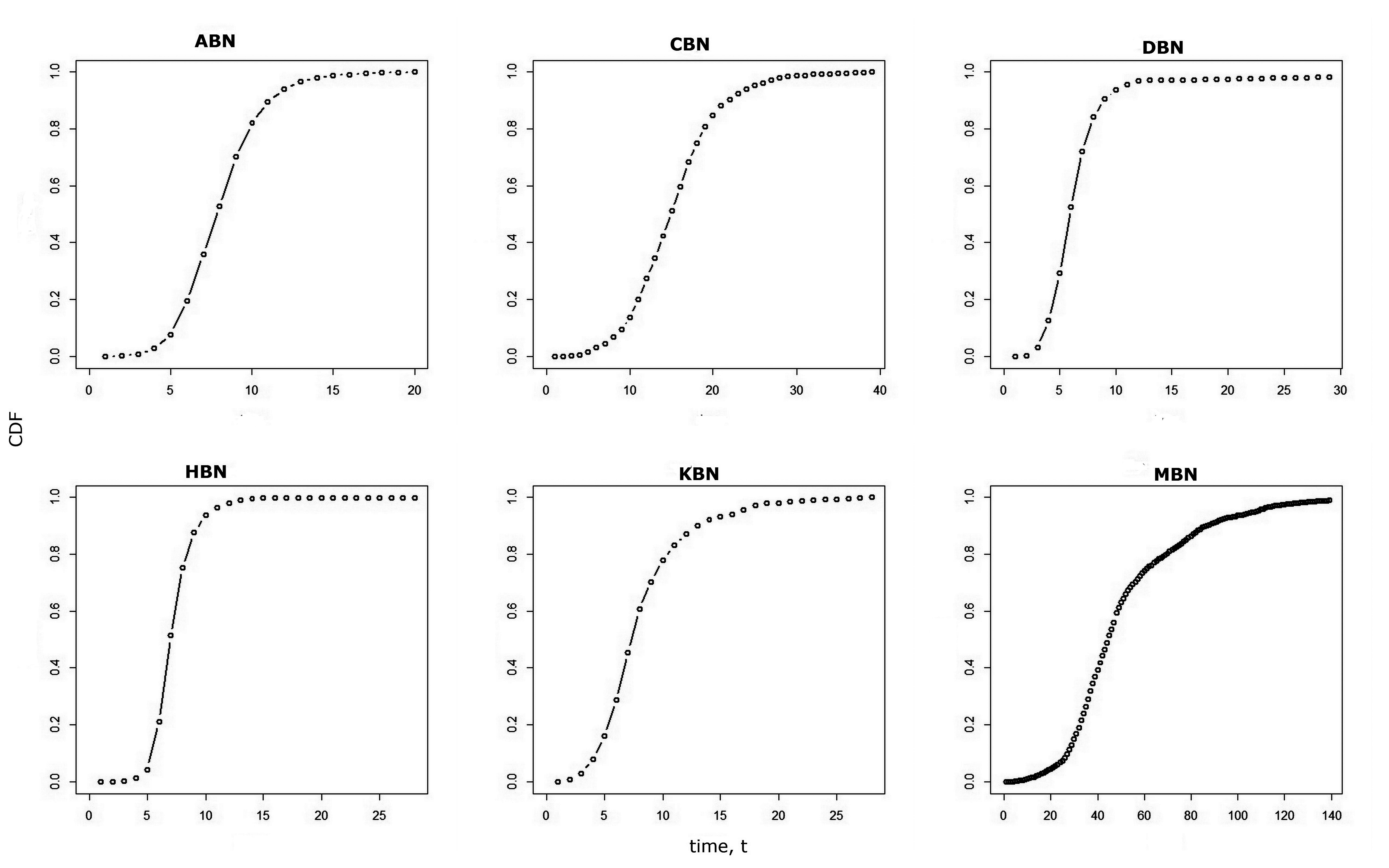}
\caption{Figure shows the SI model simulation on six different networks, with $\beta = 0.4$. The Y-axis denotes the Cumulative Distribution Function (CDF) of the infection probability of the nodes, and the X-axis represents the simulation time.}
\end{figure}

\subsubsection{Network contagion: SIR model}
In Figure 24, we plot the SIR simulation results for the different networks~\cite{chatterjee2015contagion}. The SIR curve has a typical profile because of the simultaneous decay of the infected individuals and the growth of the recovered individuals. The curve achieves a peak when the recovery rate equals the infection rate. We can clearly observe that the networks which display strong assortative behaviour (CBN and MBN) tend to have multiple peaks. The reason for the presence of multiple peaks can be explained by the fact that assortative networks tend to be hub-attractive, thus infection has multiple pathways to spread across the network, either from hub to hub, hub to node, node to hub or node to node. For weakly assortative and disassortative networks, only three among the above four possibilities exist (excluding hub to hub transmission). An infection transmitting from one hub to another hub is more likely to infect a larger number of nodes than an infection transmitted from a hub to a node. Thus, the threshold values would be achieved very early, and the presence of stochasticity in the selection of the initially infected node will induce noise in the plots followed by multiple peaks. 
\begin{figure}[hb!]
\center
\includegraphics[width=1\textwidth]{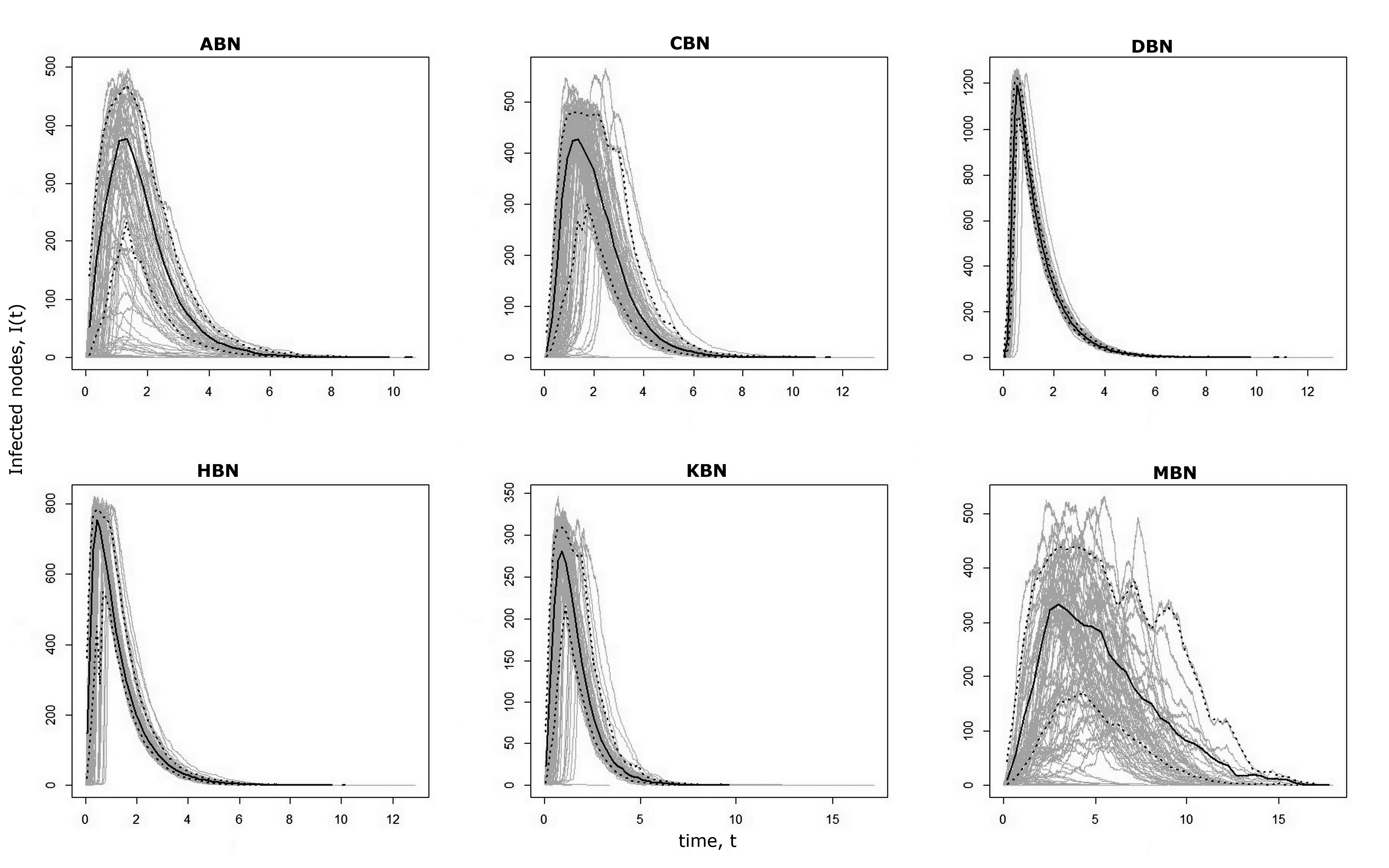}
\caption{Figure shows the SIR model simulation on the six different networks, with the parameters $\beta = 5$ and $\gamma = 1$. The Y-axis denotes the infected nodes and the X-axis represents simulation time. The curves were plotted after 100 simulations; the dark line represents the median distribution, with the dotted ones above and below as the maximum and minimum thresholds.}
\end{figure}
Similar to our previous observation, the characteristic path length $\textbf{l}_{ij}$ plays a vital role in the SIR model as well. It can be clearly observed by looking at the steepness of the plots for HBN. While ABN and DBN show similar properties, the peak and the steepness for DBN are much greater than that of ABN. This can be attributed to the fact that the average degree, $\bar{k}$ in DBN is roughly three times the average degree of ABN (see Table 1). This automatically accelerates the infection transmission rate in DBN, as each node in DBN has three times the number of choices available as compared to each node in ABN. 
\begin{figure}[hb]
\center
\includegraphics[width=1\textwidth]{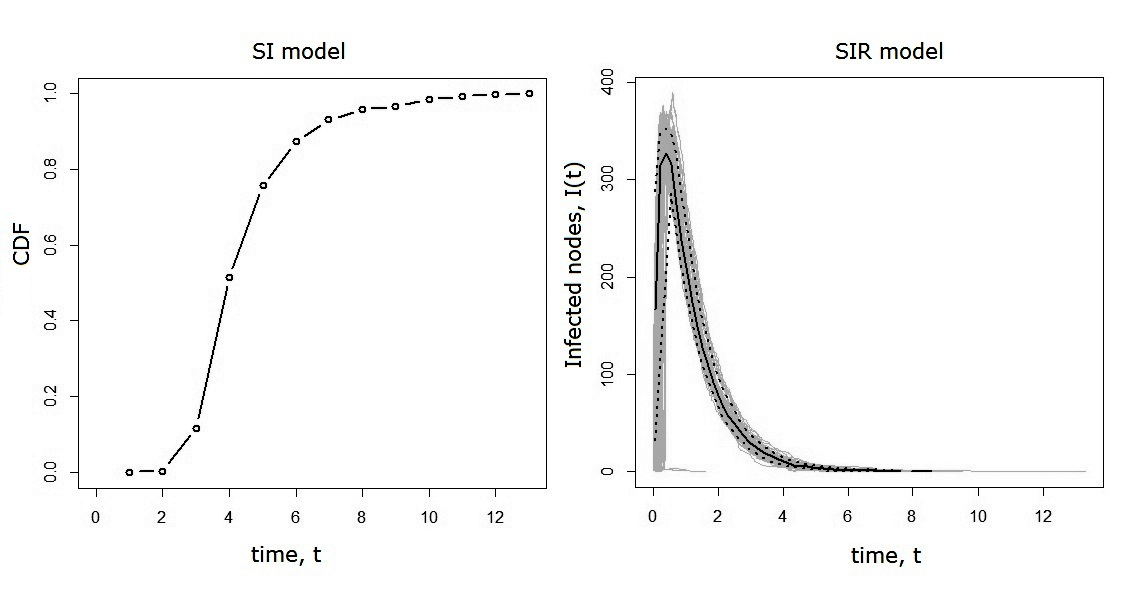}
\caption{Figure shows SI and SIR simulations on US airline network for 500 of the busiest airports~\cite{colizza2007reaction}.}
\end{figure}

\begin{table}[hb]
\centering
\begin{tabular}{|ccccc|}
\hline
             & {\bf Diffusion threshold (s)} & {\bf Epidemic threshold} & {\bf Sim. time (s)} & {\bf $\textbf{l}_{ij}$} \\ \hline
{\bf ABN}    & 10                              & 0.35                     & 1.6                 & 5.59           \\
{\bf CBN}    & 20                              & 0.42                     & 2                   & 8.73           \\
{\bf DBN}    & 8                               & 0.8                      & 1                   & 5.51           \\
{\bf HBN}    & 8                               & 0.72                     & 1                   & 3.87           \\
{\bf KBN}    & 12                              & 0.55                     & 2                   & 5.72           \\
{\bf MBN}    & 60                              & 0.15                     & 3                   & 25.69          \\
{\bf US air} & 4.5                             & 0.6                      & 1.5                 & 2.92           \\ \hline
\end{tabular}
\caption{The table outlines the $80\%$ percolation thresholds, epidemic thresholds, their corresponding simulation times and the characteristic path lengths for the various networks.}
\end{table}

In Figure 25, we simulate the SI and SIR models on the US airline network for 500 of the busiest airports ($N=500, L=2980$)~\cite{colizza2007reaction}. Statistical analysis of the network reveals scale-free degree-distribution pattern between the nodes, with the characteristic path length $\textbf{l}_{ij}=2.92$ and average degree $\bar{k} = 12$. The SI plot of the US airline network and ABN show similar pattern of growth as both of them exhibit scale-free behaviour. However, the SIR plot is similar to that of HBN due to extremely low characteristic path length and high average node degree. In Table 3, we present our findings for the networks studied in the paper. The first column represents the simulation time (in seconds) for $80\%$ percolation threshold from the SI model. In the second and third column, we present the epidemic thresholds for the various networks studied by computing the values of the plots from the SIR model (Figure 24) (as a fraction of network size) and the corresponding simulation times (in seconds) respectively. In the final column, we present the characteristic path lengths for the various networks.

\begin{figure}[t]
\center
\includegraphics[width=1\textwidth]{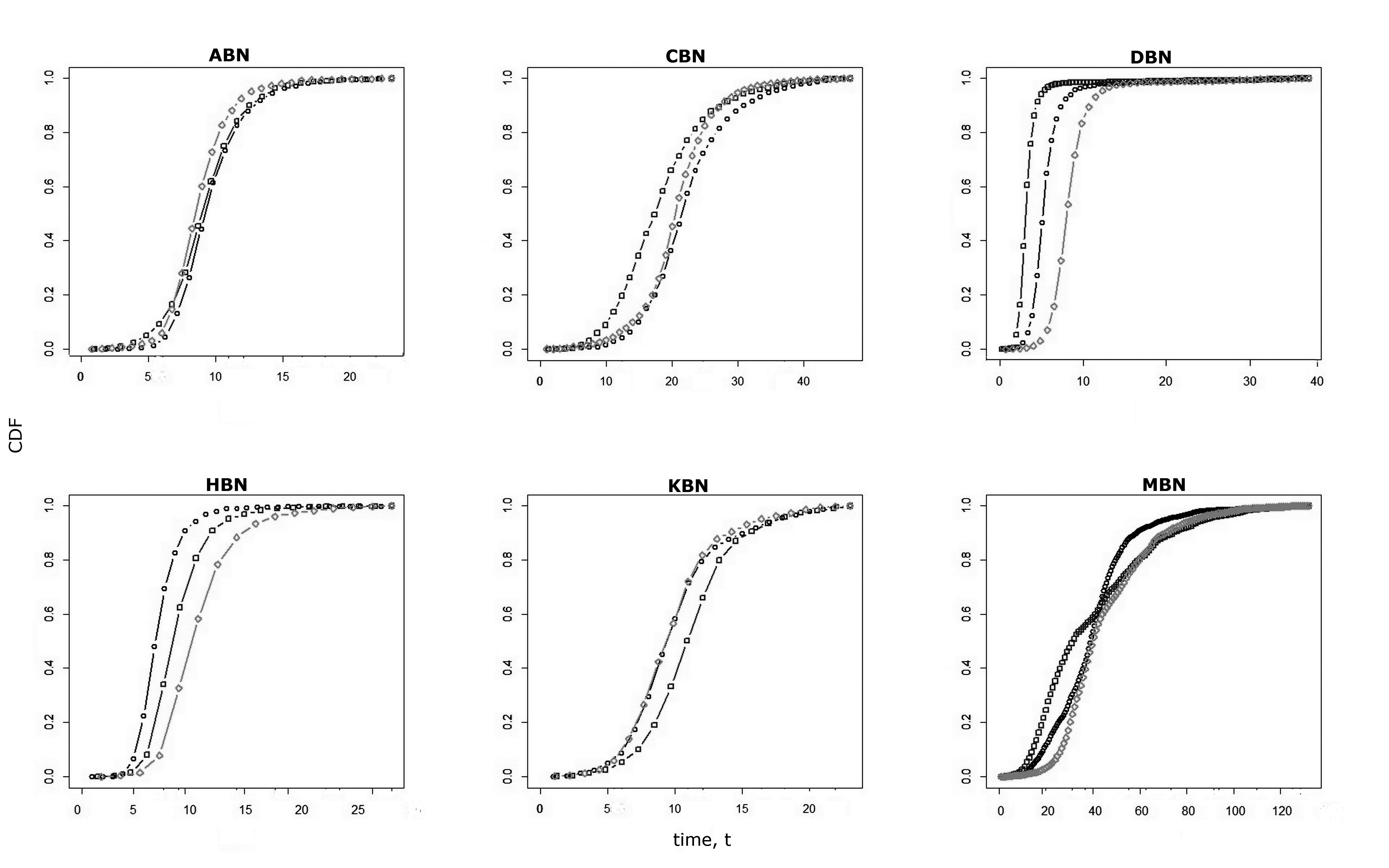}
\caption{Figure shows the SI model simulation on the six different networks after $2\%$ node removal ($\circ$ -- degree-biased, $\Box$ -- betweenness-biased and $\Diamond$ -- closeness-biased). The Y-axis denotes the CDF of the infection probability of the nodes, and the X-axis represents simulation time.}
\end{figure}

Finally, in Figure 26, we plot the variation in the rate of percolation by removing nodes from the network based upon their centralities and degrees. In transportation networks, other than the degree of a node, closeness and betweenness centralities play a crucial role. In order to capture their effects on information diffusion, we simulate the SI model on modified networks generated after directed removal of nodes. Since CBN and MBN are strongly assortative, we remove only two percent of the nodes (higher number of node removal will cause CBN and MBN to disintegrate into disconnected components). We find that the removal of nodes does not significantly affect the diffusion in ABN. However for CBN, DBN and HBN, we observe that when nodes are removed based upon their closeness centrality, the diffusion curve shifts towards right, thus signifying a delay in the diffusion process. This can be explained due to the fact that the removal of nodes based upon closeness centrality has a direct effect on the characteristic path length. A node with high closeness allows every other node in the network to be reached along the shortest paths. The removal of such a node affects/delays diffusion until the next central node is encountered. For MBN, we observe that degree-biased removal causes the diffusion rate to increase steeply, signifying the presence of redundant nodes that simply increase the characteristic path length of the network. A removal of $2\%$ of such nodes causes the diffusion to improve significantly, as can be compared from the simulation times recorded in Figure 23. 

\section{Scope of future works}
The present work places before us numerous questions of both academic as well as practical pursuit. One important practical scope of this work is in the planning of large-scale transportation networks for the future. It would be interesting to see the functionality of those networks which are planned using network science tools and techniques. Another question of practical importance lies in optimizing these networks for efficient transportation and communication purposes. One important finding that could be of practical interest is the strong positive correlation between network assortativity and characteristic path length: will it be more efficient to travel on longer routes to reach hub nodes and from there to other parts of the city, or to travel though shorter routes, not reach the hub nodes but travel to other parts of the city through intermediate nodes?

From academic pursuit, there are numerous scopes for in-depth analysis based upon this study. The current study takes into account one subset of the large-scale public transit networks. It would be interesting to do an integrated study involving other modes of transport as well. The availability of high quality geo-located data will help in actually identifying redundant nodes in the network, thus making the network more efficient. Since transportation plays an important role in the economic development of a city, the present study can be extended to incorporate other networks as well, such as road networks, supply chain networks and economic networks. A holistic approach to all these networks will help us in understanding each layer of society's complexity. 

\begin{figure}[hb!]
\center
\includegraphics[width=1\textwidth]{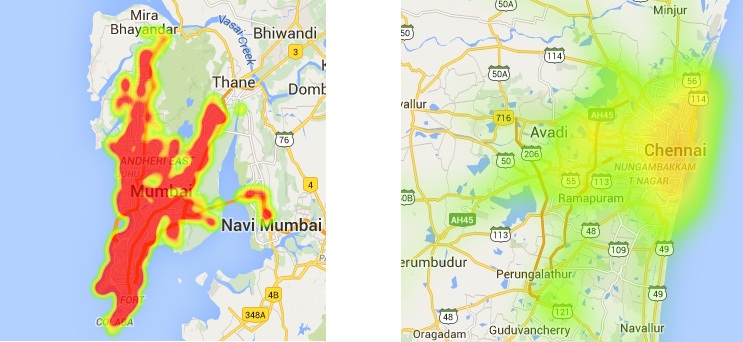}
\caption{Figure shows the distribution of bus stops in Mumbai (left) and Chennai (right).}
\end{figure}

\section{Discussion and Conclusion}
In this study, we analyzed the statistical properties of the bus routes of six Indian cities, namely Ahmedabad, Chennai, Delhi, Hyderabad, Kolkata and Mumbai. Our analysis suggests that the bus networks show a wide spectrum of topological structure from power-law to exponential, with varying magnitude of the power-law exponent $\gamma$. We also observe that these networks show small-world behaviour in terms of either the placement of the bus stops ($L$-space) or in terms of transfers ($P$-space). For example, CBN and MBN do not show the small-world property in $L$-space. They, however, do show the small-world property in terms of transfers, as the majority of places can be visited by making as little as 2 to 3 bus changes. The redundancy in the network structure, as seen from the variation in $\textbf{l}_{ij}$ and the presence of exponential cutoffs in the degree-distribution plots, validate our findings regarding the randomness associated with the growth and evolution of the bus networks with time. Recently, Wang et al. simulated exponential growth models for networks, with growth and adjacent node attachment as underlying processes. The growth equation according to their model is as follows: $\frac{\partial P(k)}{\partial t} + P(k, t) = \frac{t\partial P(k, t)}{\partial t} = P(k-1, t) - P(k, t) + \delta_{k, 1}$, where $P(k, s = t, t > 2) = \delta_{k,1}$. The variable $s = 1, 2, 3...$ marks the vertices, and $P(k, s, t)$ represents the probability that a vertex $s$ has a degree $k$ at time $t$. The stationary degree distribution, $P(k)=P(k, t\rightarrow\infty)$ is given as $2P(k) - P(k-1) = \delta_{k,1}$. In the continuum limit, the above equation approaches the form $\frac{dP(k)}{dk} = - P(k)$, giving rise to the decay equation $P(k) \sim \exp(-\lambda k)$, with $\lambda = \bar{k}^{-1}$~\cite{deng2011exponential}.

Our findings on the weak correlations between degree and centrality plots and self-similar structures of sub-networks motivate us to investigate the fractal nature of these networks. The degree-degree correlation matrix gives a rough idea on the degree-assortativity of these networks. Strong assortativity in networks relate to strong connections between hubs, thus hubs tend to come in-between short-range connections. Whereas, in disassortative mixing we observe hub-repulsion. The presence of hub-inter connectedness causes the characteristic path lengths of the networks to increase and also the fractal nature to diminish. Presence of repelling hubs generate fractal topology in complex networks. An interesting aspect of the fractal topology is that they generate local small-worlds in well-defined communities. From a transportation perspective, it would be beneficial to have local small-worlds connected to a central core, as such a structure will reduce the characteristic path length and make the network efficient. 

The high values of characteristic path lengths for CBN and MBN can be attributed to the geographical structures of the two cities, Chennai and Mumbai. The routes in these two cities are exceptionally longer because the BTNs in these two cities have evolved more in a linear fashion. The reason for this may lie in the geographical limitation imposed by the presence of a water body on one side (see Figure 27). Also, most of our results on Ahmedabad BRTS are inconclusive as the dataset is small. At present, the BRTS operates across only 13 routes with 129 bus stops. Therefore, it would be interesting to see the evolution of this network in future.

Finally, we simulate SI and SIR dynamical processes on these networks in $L$-space. Since experiments with epidemic outbreaks in a population (or a network) is not a viable option, we resort to mathematical modelling to understand diffusion of information and contagion spreading. We therefore study the effect of percolation and epidemic spreading on these networks using SI and SIR epidemic models through numerical simulations. While it is observed that the characteristic path length plays a crucial role in information diffusion and epidemic spreading, several other network metrics also play important roles. Their importance is however restricted to their relative contribution to the topological structures of the networks. Small-world property, while an extremely desirable property in transportation networks, is highly subjective in its role of information diffusion, solely due to the diffusing entity. 

Finally, bus networks form a specific class of complex networks that grow and evolve over physically constrained spatial networks. Interesting in this regard is the city of Ahmedabad and the ABN. Statistical analysis of road networks (by considering intersections as nodes and roads as links) has shown that the topological structure of the road networks in the city of Ahmedabad exhibits a scale-free degree distribution with $\gamma = 2.5$ and $l_{ij} = 5.20$, which is very much similar to ABN~\cite{porta2006network}. Road intersections are usually separated by a distance which is geographically much smaller as compared to the distance between bus stops. Therefore, our results emphasize the fact that transportation undoubtedly brings the world closer.

\end{document}